\documentclass[fleqn,usenatbib]{mnras}

\usepackage{newtxtext,newtxmath}

\usepackage[T1]{fontenc}
\usepackage{ae,aecompl}

\usepackage{graphicx}	
\usepackage{amsmath}	

\usepackage{multicol}        
\usepackage{bm}		
\usepackage{pdflscape}	
\usepackage{multicol}
\usepackage{graphicx}
\usepackage{caption}
\usepackage[utf8]{inputenc}
\usepackage{booktabs, caption, makecell}

\usepackage{threeparttable}
\usepackage[table]{xcolor}
\usepackage{tabulary}
\usepackage{siunitx}
\usepackage{hyperref}
\usepackage{natbib}

\title[X-ray observations of M62]{Multi-Epoch X-ray Imaging of Globular Cluster M62 with {\it Chandra}}

\author[K. Oh et al.]{Kwangmin Oh,$^{1}$\thanks{E-mail: min8582046@gmail.com}
C. Y. Hui,$^{2}$\thanks{E-mail: huichungyue@gmail.com}
K. L. Li$^{3}$
and A. K. H. Kong$^{3}$
\\
$^{1}$Department of Space Science and Geology, Chungnam National University, Daejeon 34134, Korea\\
$^{2}$Department of Astronomy and Space Science, Chungnam National University, Daejeon 34134, Korea\\
$^{3}$Institute of Astronomy, National Tsing Hua University, Hsinchu, 30013, Taiwan\\
}

\date{Accepted XXX. Received YYY; in original form ZZZ}

\pubyear{2015}

\begin{document}
\label{firstpage}
\pagerange{\pageref{firstpage}--\pageref{lastpage}}
\maketitle

\begin{abstract}
Using archival spectral-imaging data with a total exposure of $\sim144$~ks obtained by {\it Chandra}, 43 X-ray sources are detected within the half-light radius of globular cluster M62 (NGC6266). Based on the X-ray colour-luminosity diagram or the positional coincidences with known sources, we have classified these sources into different groups of compact binaries including cataclysmic variable (CV), quiescent low mass X-ray binary (qLMXB), millisecond pulsar (MSP) and black hole (BH). Candidates of the X-ray counterparts of 12 CVs, 4 qLMXBs, 2 MSPs and 1 BH are identified in our analysis. The data used in our analysis consist of two frames separated by 12 years, which enable us to search for the long-term variability as well as the short-term X-ray flux variability within each observation window. Evidence for the short-term variability and long-term variability have been found in 7 and 12 sources respectively. For a number of bright sources with X-ray luminosities $L_{x}\gtrsim 10^{32}$~erg/s, we have characterized their spectral properties in further details. 
By comparing the X-ray population in M62 with those in several other prototypical globular clusters, we found the proportion of bright sources is larger in M62 which can possibly be a result of their active dynamical formation processes. 

\end{abstract}




\begin{figure*}
\includegraphics[width=\textwidth]{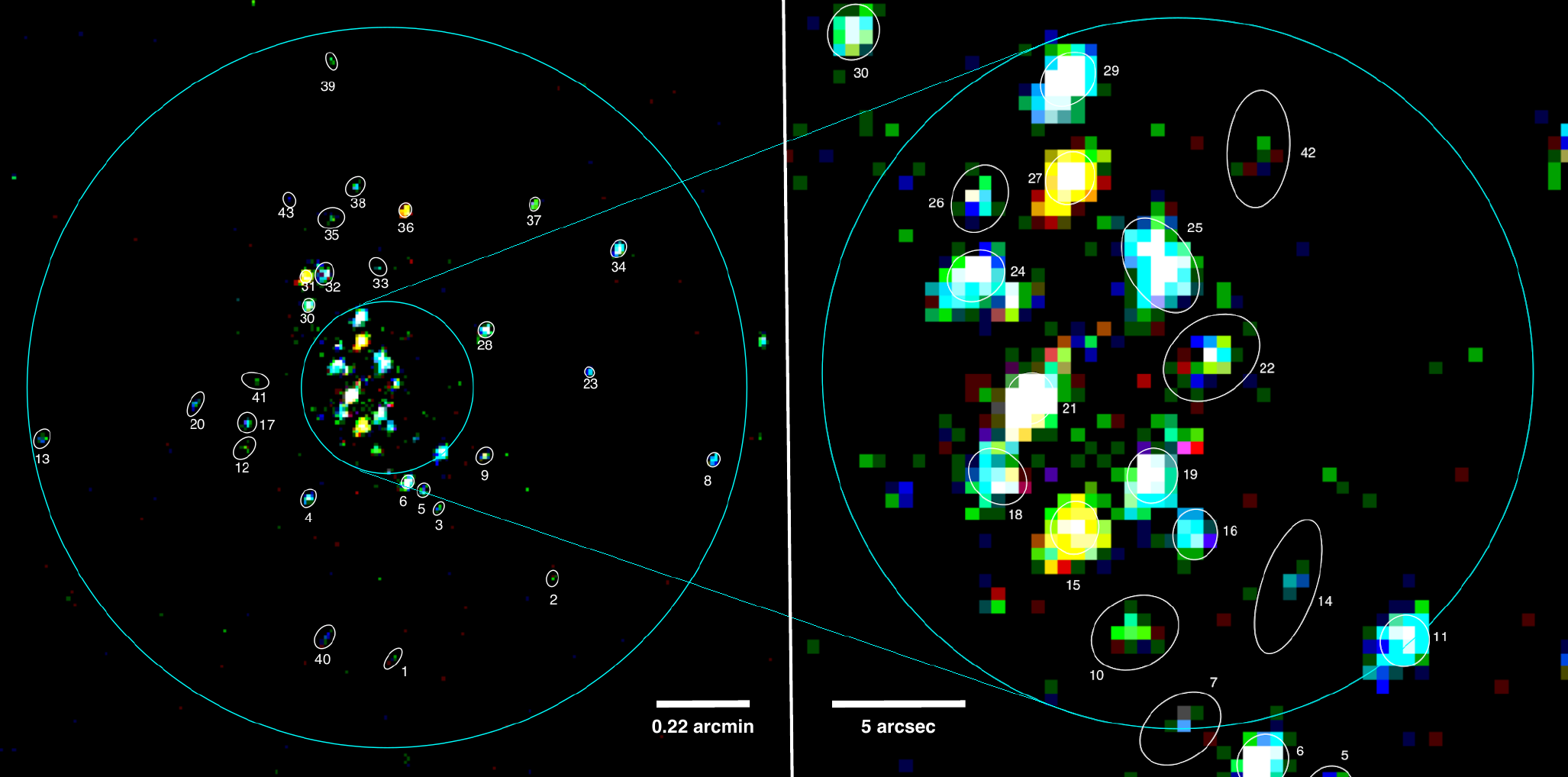}
\caption{X-ray colour images of M62 with both {\it Chandra} ACIS observations merged (red:0.3-1.0keV, green:1.0-2.0keV and blue:2.0-7.0keV). ({\bf Left panel:}) The field covers the half-light radius is displayed. The large and small blue circles illustrate the half-light radius (0.92 arcmin) and core radius (0.22 arcmin) of M62 respectively. 43 sources were detected from the merged data. The numbering scheme of these source is in accordance with that in Table~1, in which the X-ray properties of these sources are summarized. ({\bf Right panel:}) The close-up X-ray view of the core region of M62.}	
\label{fig:example_figure}
\vspace{5mm}
\end{figure*}

\section{Introduction}
Globular clusters (GCs) are the densest stellar systems which typically contain $\sim{10^{4}}$ - ${10^{7}}$ stars in a region with a radius of $\lesssim100$~pc. Owing to the high stellar density in their central regions, various classes of compact binaries such as cataclysmic variables (CVs), low-mass X-ray binaries (LMXBs), millisecond pulsars (MSPs) can be formed dynamically \citep{pooley2003,pooley2006,hui2010}. The binary population in a GC can play an important role in determining the equilibrium state of the cluster and hence its evolution  \citep[e.g.][]{fregeau2008}. Therefore, revealing the populations of compact binaries in different GCs can lead us to a deeper understanding of how these clusters evolve.

Since most of the compact binaries are X-ray emitters, X-ray imagings of GCs can provide an effective way to uncover the binary populations residing in their central regions. Since the launch of {\it Chandra}, a large number of X-ray sources have been resolved from different GCs \citep[e.g.][]{heinke2005}.
This allowed some meaningful statistical analysis to be performed \citep{pooley2003,pooley2006}. For example, through investigating the relation between the X-ray population in a GC and the intrinsic properties of the cluster, \cite{pooley2003} have found a strong correlation between the population of LMXBs and the stellar encounter rate $\Gamma_{c}$ in GCs which suggests their dynamical origin.

M62 (NGC 6266) is among the GCs which have the highest stellar encounter rate \citep[cf. Figure 2 in][]{pooley2003}. Its cluster centre is located at 
RA=$17^{\rm h}01^{\rm m}12.80^{\rm s}$  Dec=$-30^{\circ}06^{'}49.4{"}$ (J2000) \citep{harris}. Its core radius and half-light radius are $\theta_{c}=0.22$~arcmin and $\theta_{h}=0.92$~arcmin, respectively (\cite{harris}). At a distance of 6.8~kpc, these correspond to a physical size of $r_{c}=0.4$~pc and $r_{h}=1.8$~pc. 
The foreground reddening is found to be $E(B-V)=0.47$ \citep{harris}. 
 
Radio pulsar surveys have so far detected six MSPs in M62 \citep{lynch}. Five of them are located within $\theta_{c}$. All these MSPs are in
binaries including two black-widows and one redback (See \cite{hui2019} for an updated review on black-widow and redback MSPs). As limited by the sensitivity of the current radio pulsar searches, its intrinsic MSP population can possibly be larger. {\it Fermi} Gamma-ray Space Telescope has detected $\gamma$-ray emission from M62 in MeV-GeV regime \citep{abdo}. If one assumes the $\gamma$-rays are originated from the collective contribution of the magnetospheric emission of the entire MSP population in the cluster, $\sim80$ MSPs are expected to be found in M62 \citep{abdo}. 

On the other hand, a deep radio continuum imaging with Very Large Array (VLA) has discovered a source with a flat radio spectrum, which has a spectral index of $\alpha=-0.24$ \citep{chomiuk}. Such property makes this source distinct from a radio pulsar which typically has a steep radio spectrum. Furthermore, X-ray and optical counterparts of this radio source have also been identified \citep{chomiuk}. Since its X-ray, radio and optical properties are similar to the well-studied V404 Cygni in quiescence, this source has been suggested to be a stellar mass black hole candidate \citep{chomiuk}. 

While the high resolution X-ray imaging data of M62 obtained by {\it Chandra} have been briefly utilized in several previous published works \citep[e.g.][]{pooley2003,hui2009,chomiuk}, there is no systematic study for characterizing the entire population of the X-ray sources. We also noted that M62 has been observed by {\it Chandra} twice with an epoch separation of $\sim12$~years. This allows us to search for the X-ray variability which is typical behaviour of compact binaries. In this paper, we report a systematic analysis of the multi-epoch {\it Chandra} observations of M62 so as to characterize its X-ray sources in detail. In Section~2, we describe the observations and the adopted methods in processing the data. In Section~3, we present a systematic analysis for characterizing the basic X-ray properties of all the sources detected within the half-light radius and an attempt to classify them according to their X-ray properties. The results from the searches of temporal variability are summarized in Section~4. For the bright sources, we have further examined their spectral properties with detailed analysis and the results are given in Section 5. Lastly, we summarize our results and discuss their implications in Section 6.

\section{observation and data reduction}
M62 has been observed by  Advanced CCD Imaging Spectrometer (ACIS) on board {\it Chandra} for the exposures of $\sim62$~ks and $\sim82$~ks during 2002 May 12-13 (Obs IDs 2677l; hereafter obs1) and 2014 May 05-06 (Obs IDs 15761; hereafter obs2) respectively. In both observations, the centre of M62 is on ACIS-S3 which is one of the two back-illuminated CCDs on board. We note that the cluster centre of M62 has an off-axis angle of 2.39 arcmin in obs1 and 0.18 arcmin in obs2. 

Using the script {\tt\string chandra\_repro} in CIAO (version 4.9), we have re-processed both data with the updated calibration (CALDB version 4.7.7). 
To facilitate source detection with high positional accuracy, we have applied the sub-pixel event repositioning in reprocessing the data.
The task generated bad pixel files and level=2 event files which were used for the following analyses.We restricted all the analyses in an energy band of $0.3-7$~keV.

\section{X-ray sources within half light radius}
\subsection{Source detection}

We used the CIAO task {\tt\string wavdetect} to perform source detection in the merged data as well as individual observations. We utilized the script {\tt\string merge\_obs} for merging the event files of two observations, which can reproject the same tangent point of two event files, merging them into a single event file. We have inspected the result by visually examining the merged data and we do not find any anomaly in the image of point sources. At the same time, the script also creates exposure maps for each observation and enables us to generate the exposure-corrected images for correcting the vignetting effect by using the script {\tt\string fluximage}.

For determining the kernel parameters, we adopted a point spread function (PSF) map at 1.5 keV and chose an enclosed counts fraction (ECF) of 39\% after experimenting with the data so as to resolve the crowded sources in the core region. Pixel scales have been fixed at 1.0, 1.4, 2.0, 2.8. For the significant threshold for source detection, we adopt a value of 10$^{-6}$ which corresponds to $\sim1$ false alarm for running the detection on a 1024 pixels $\times$ 1024 pixels image. In this work, we only consider a detection with a signal-to-noise ratio $>3$ as a genuine source. 

In total, 43 X-ray sources have been detected within the half-light radius from the merged data. 16 of them are located within core radius. The results are summarized in Table~1. The $1\sigma$ statistical errors of the X-ray positions of all these sources are calculated by the net counts-weighted variances around the centroid. We would like to emphasize the absolute astrometric uncertainty by {\it Chandra}, which has a 90\% uncertainty circle with a radius of $\sim0.8$~arcsec\footnote{https://cxc.harvard.edu/cal/ASPECT/celmon/}, is not included in columns 4 and 5 in Table~1.

In Figure~1 we show the colour-coded images (Red: 0.3-1~keV; Green 1-2~keV; Blue 2-7~keV) which cover the half-light radius (left panel) and the core radius (right panel). All the sources detected in our work are highlighted by the green ellipses with the numbering scheme consistent with that given in Table~1. 

We also have examined if these sources can be detected in individual observations. In the column~7 of Table~1, we have added the flags ``Y/N" to indicate whether the source is detected/undetected for each observation respectively. 7 of them cannot be detected in obs1 and 8 of them cannot be detected in obs2, which indicate the possible long-term X-ray flux variabilities of these sources (see Section 4.1).

The detection significance of source s2 is above $3\sigma$ in the merged data but it is found to be below our predefined detection threshold in both individual observations. Therefore, its properties reported in this work is based on the merged data.

\subsection{Flux estimation}
We proceeded to estimate the unabsorbed energy flux of all these sources in each observation. For correcting the variation of instrumental responses, we have extracted the source and background spectra from the sources detected in each individual observation and generate the corresponding response files by using the CIAO tool {\tt specextract}. Each spectrum is fitted by an absorbed power-law model with the photon index fixed at $\Gamma=1.7$. With the column absorption fixed at $N_{H}=3.2\times10^{21}$~cm$^{-2}$ as inferred by the foreground reddening of M62 $E(B-V)=0.47$ \citep{guver}, we estimate the unabsorbed energy fluxes and their $1\sigma$ uncertainties of each source with the aid of convolution model {\tt\string cflux} in the spectral fitting package {\tt\string XSPEC}. The results are summarized in the column 8 of Table~1.

\subsection{X-ray colour-luminosity diagram}

\begin{figure*}
	\includegraphics[width=17cm]{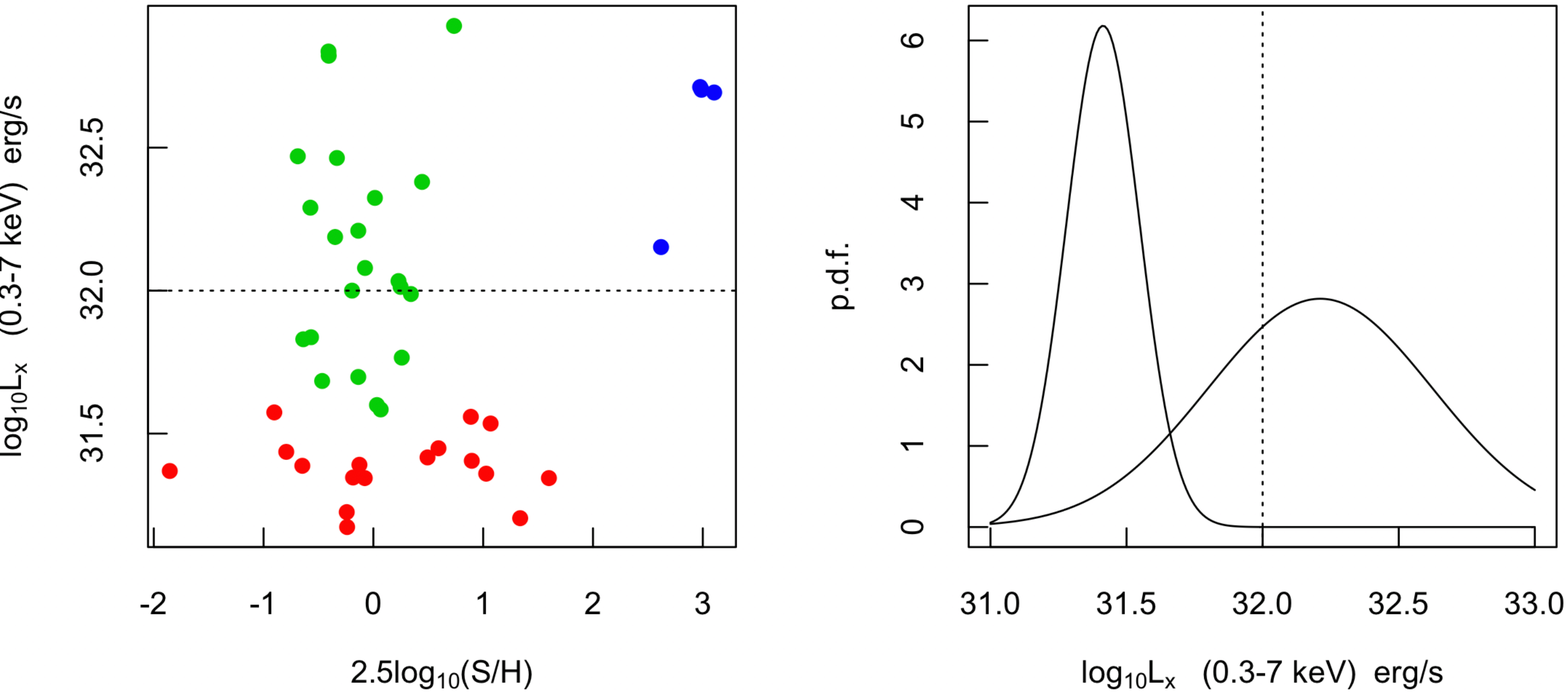}
\vspace{-1cm}
    \caption{({\bf Left panel:}) Unsupervised classification of the X-ray colour-luminosity distributions of M62 sources by the method of 2-dimensional Gaussian mixture model (GMM). The population can be divided into three clusters which are represented by the symbols of different colours. ({\bf Right panel:}) Marginalized Gaussian distributions of X-ray luminosities given by GMM. The overlap of these two distribution in the range $L_{x}\lesssim5\times10^{31}$~erg/s can lead to misclassification. For a clearer identification solely based on X-ray data, we only consider the possible nature of the sources with $L_{x}>10^{32}$~erg/s in this work. This predefined threshold is illustrated by the dashed line in both panels.} 
    \label{fig:example_figure}
\end{figure*}

We further attempt to classify the X-ray sources in M62 by constructing the X-ray colour-luminosity diagram. For the calculating the luminosity of each sources, we multiply the unabsorbed fluxes by 4$\pi$$d{^{2}}$, where $d$ is 6.8 kpc \citep{harris}. The limiting luminosity for a detection threshold $>3\sigma$ of obs1, obs2 and the merged data are found to be $L^{\rm lim}_{\rm obs1}\gtrsim10^{31}$~erg~s$^{-1}$, $L^{\rm lim}_{\rm obs2}\gtrsim5\times10^{30}$~erg~s$^{-1}$ and $L^{\rm lim}_{\rm merged}\gtrsim3\times10^{30}$~erg~s$^{-1}$ respectively. 
In this work, we define the X-ray colour as 
$2.5\log\left(S/H\right)$ where $S$ and $H$ are the net counts in the soft band (0.3-2~keV) and hard band (2-7~keV) respectively. 

Visual inspection on source distribution in the X-ray colour-luminosity diagram of M62 suggests there are possible distinct groupings. For quantifying this, we have analysed the distribution with the Gaussian Mixture Model (GMM) algorithm. Based on the Bayesian information criteria (BIC), the source distribution in the X-ray colour-luminosity diagram of M62 requires three 2-D Gaussian components to model. In the left panel of Figure~2, these three different groups are represented by symbols of different colours. 

While the group of luminous soft X-ray sources (i.e. the group on the upper-right corner of the colour-luminosity diagram) is well separated from the others, the other two groups are overlapped in the luminosity distribution (see the right panel in Figure~2). In order to avoid misclassifications, we apply a cut at ${10^{32}}$~{erg~s$^{-1}$} so that the luminous hard X-ray sources (i.e. the group on the upper-left of the colour-luminosity diagram) can be cleanly separated from the less luminous hard sources (i.e. the group on the lower-left of the colour-luminosity diagram). 

By comparing these groupings with X-ray colour-luminosity diagrams of the other GCs \citep[e.g.][]{heinke2005,heinke2014,henleywillis}, these groups can be associated with different classes of compact binaries. As most of the CVs in GCs are characterized by their hard X-rays and high luminosities \citep[e.g.][ and references therein]{heinke2005}, this suggests that the upper-left group in our colour-luminosity diagram is likely to CVs. With the aforementioned consideration of avoiding misclassifications, we consider the sources with luminosities $L_{x}>{10^{32}}$~{erg~s$^{-1}$} and colours $2.5\log\left(S/H\right)<1$ as promising CV candidates. In this work, 12 sources are considered as CV candidates. For classifying the less luminous hard X-ray sources, optical observations will be required to provide additional information which is out of scoop in this work. 

For the group of soft and luminous X-ray sources (i.e. $L_{x}>{10^{32}}$~{erg~s$^{-1}$}, $2.5\log\left(S/H\right)>2$), their locations in the X-ray colour-luminosity diagram are consistent with those of qLMXBs in other GCs \citep{heinke2005}. qLMXBs are typically characterized by the soft blackbody-like spectral component \citep{heinke2014}. Among our source list in Table~1, four of them belong to this group and hence they are considered as qLMXB candidates. \cite{arash2015} have also studied the qLMXBs in M62 and suggested four qLMXB candidates from their analysis. The qLMXB candidates selected by our scheme are identical with those reported by \cite{arash2015}.  

In Figure~3, we have plotted the X-ray colour-luminosity diagram with all the sources tagged by their source numbers in Table~1. $1\sigma$ error bars of both X-ray luminosites and colours are also shown in this plot. The candidates of CV and qLMXB suggested by the aforementioned procedures are displayed as the blue triangles and red diamonds in Figure~3 respectively. For the sources with $L_{x}<{10^{32}}$~{erg~s$^{-1}$} and $2.5\log\left(S/H\right)<2$ as aforementioned, they can be a mixture of different classes of objects. In order to classify them, multiwavelength data (e.g. optical) are required which is out of scoop in this work. 

Apart from recognizing the clusters of data points in the X-ray colour-luminosity diagram, one can also identify the nature of the X-ray sources by the positional coincidence with the other known objects in the GC. For the six MSPs detected in M62, the radio timing positions of PSR J1701-3006B and PSR J1701-3006C are found to be consistent with the X-ray sources s22 and s10 respectively within the tolerence of their positional uncertainties. PSR J1701-3006B is a redback MSP which has an orbital period of $P_{b}\sim0.14$~days and a companion with a mass of $>0.12$~$M_{\odot}$ (\cite{lynch}; Tab.~2 in \cite{hui2019}). 

On the other hand, the radio position of a BH candidate \citep{chomiuk} is consistent with that of the brightest X-ray source detected in M62, i.e. s21. Its X-ray colour appears to be softer than the group of CV candidates but harder than that of the qLMXB candidates. 

Besides s21, source s43 is another possible outlier. The data only allow us to place an upper-limit of its X-ray colour (see Figure~3) which indicates it is the hardest source found in this field. And yet it is very faint (see Figure~1). This leads us to speculate that this source is likely to be a background AGN. As estimated by \cite{pooley2003}, there are approximately 2 to 3 background sources within M62's half-light radius. Apart from s43, there might still be one or two objects in our source list awaited to be identified as background objects.

In the column 11 of Table~1, the possible nature of the sources as identified in this section are summarized.

\begin{figure*}
	\includegraphics[width=17cm]{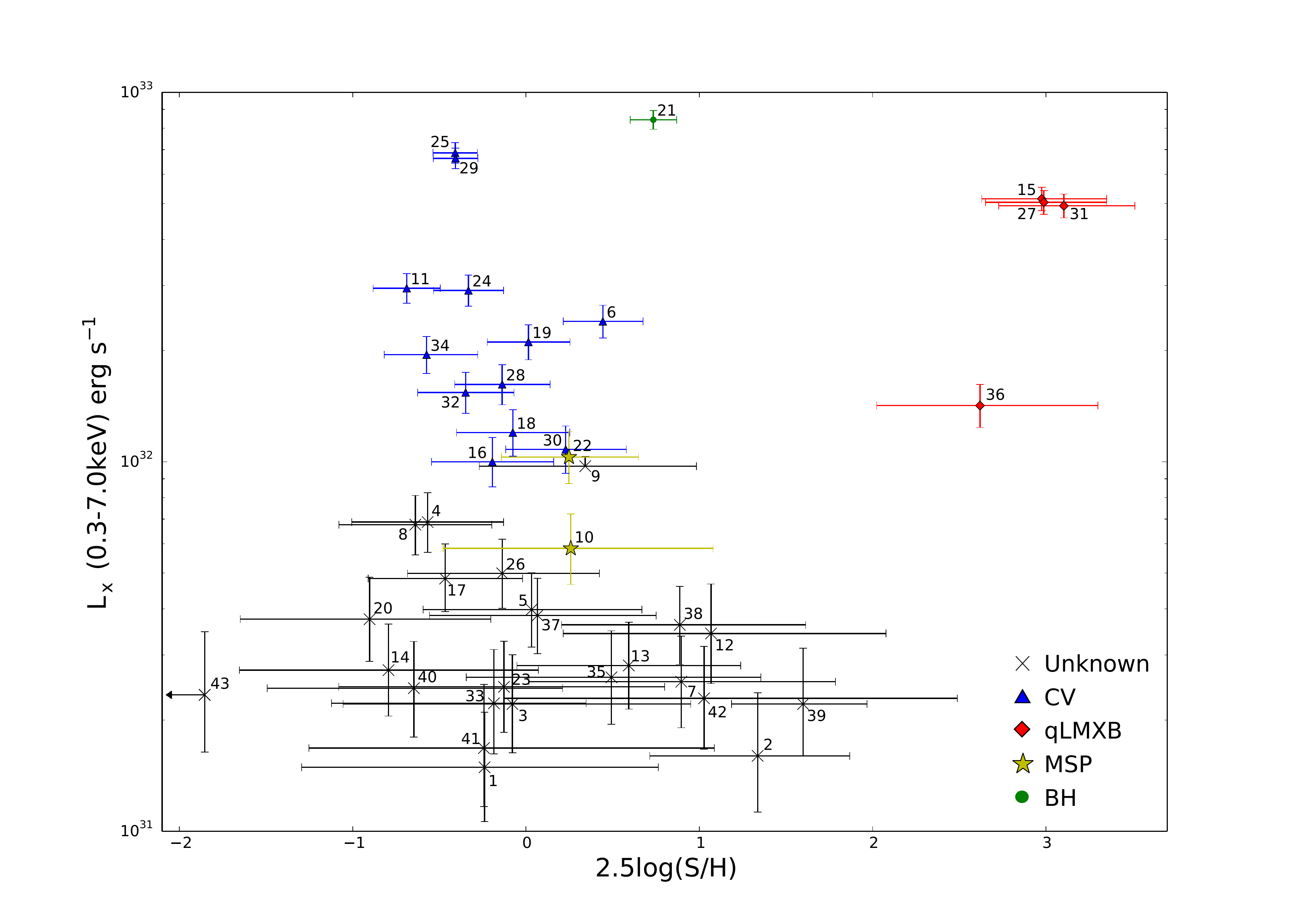}
    \caption{X-ray colour-luminosity diagram of 43 detected X-ray sources within half-light radius of M62 obtained from the merged data. The plotted error bars show $1\sigma$ uncertainties of both colour and luminosity. The possible natures of these sources identified in this work are represented by different symbols. } 
\end{figure*}

    \begin{table*}
        \caption{X-ray properties of X-ray sources within the half-light radius of M62.}
         \label{tab:landscape}
         \begin{tabular}{ccccccccccc}
 \hline
{Source} & {R.A.}$^{(1)}$ & {Dec.}$^{(1)}$ & {$\delta$$_{\rm RA}$}$^{(2)}$ & {$\delta$$_{\rm Dec}$}$^{(2)}$ & {S/N}$^{(3)}$ & {obs1 / obs2}$^{(4)}$ & {$F_{\rm obs1}$ / $F_{\rm obs2}$}$^{(5)}$ & {\it{$\theta_{\rm centre}$}$^{(6)}$} & {Variability}$^{(7)}$ & {ID}$^{(8)}$ 
\\
 &  &  & {(arcsec)} & {(arcsec)} &  & {(Detection)} & {($10^{-15}$erg/cm$^2$/s)} & {(arcsec)} & {(S/L)} &  
\\
\hline
$s01$ &  17 01 12.727 &  -30 07 31.017&         0.127   &0.128  &6.05    &Y/N   & 2.07	 $^{+1.29}_{-0.96}$ / -  &41.628 & L* & \\
$s02$ &  17 01 10.846 &  -30 07 18.676&         0.097   &0.118  &4.62    &N/N   & - / -  &38.762 & -  & \\
$s03$ &  17 01 12.188 &  -30 07 07.947&         0.062   &0.066  &7.95    &Y/Y   & 3.48	 $^{+1.65}_{-1.26}$ /1.88	 $^{+1.27}_{-0.89}$  &20.174 & -  & \\
$s04$ &  17 01 13.728 &  -30 07 06.384&         0.044   &0.045  &29.96   &Y/Y   & 10.21	 $^{+2.58}_{-2.21}$ /9.73	 $^{+2.42}_{-2.08}$  &20.820 & -  & \\
$s05$ &  17 01 12.365 &  -30 07 05.163&         0.054   &0.053  &11.08   &N/Y   & - /6.69	 $^{+2.03}_{-1.69}$  &16.740 & S(1), L*  & \\
$s06$ &  17 01 12.554 &  -30 07 03.868&         0.018   &0.017  &56.94   &Y/Y   & 36.65	 $^{+4.65}_{-4.29}$ /40.53	 $^{+4.51}_{-4.20}$  &14.815 & S(1) & CV\\
$s07$ &  17 01 12.790 &  -30 07 02.640&         0.094   &0.074  &6.74    &N/Y   & - /2.18	 $^{+1.29}_{-0.94}$  &13.240 & L*  & \\
$s08$ &  17 01 10.893 &  -30 07 00.481&         0.042   &0.042  &30.7    &Y/Y   & 6.17	 $^{+2.02}_{-1.66}$ /12.51	 $^{+2.77}_{-2.41}$  &51.352 & -  & \\
$s09$ &  17 01 11.644 &  -30 06 59.990&         0.058   &0.054  &17.42   &Y/Y   & 6.74	 $^{+2.13}_{-1.78}$ /23.82	 $^{+0.35}_{-1.41}$  &18.301 & -  & \\
$s10$ &  17 01 12.920 &  -30 06 59.085&         0.076   &0.57   &9.14    &Y/Y   & 6.93	 $^{+2.32}_{-1.93}$ /8.82	 $^{+2.70}_{-2.24}$  &9.811  & -  & MSP\\
$s11$ &  17 01 12.147 &  -30 06 59.389&         0.017   &0.015  &77.16   &Y/Y   & 46.62	 $^{+5.35}_{-4.96}$ /49.32	 $^{+4.99}_{-4.67}$  &13.089 & -  & CV\\
$s12$ &  17 01 14.483 &  -30 06 58.700&         0.121   &0.102  &7.64    &Y/Y   & 5.46	 $^{+3.15}_{-2.33}$ /2.82	 $^{+1.55}_{-1.15}$  &23.735 & -  & \\
$s13$ &  17 01 16.878 &  -30 06 57.237&         0.078   &0.08   &11.42   &Y/Y   & 1.87	 $^{+1.38}_{-0.93}$ /4.76	 $^{+1.72}_{-1.42}$  &53.498 & -  & \\
$s14$ &  17 01 12.481 &  -30 06 57.376&         0.082   &0.14   &6.06    &Y/N   & 5.47	 $^{+2.04}_{-1.64}$ / -  &8.984  & S(1,2), L* & \\
$s15$ &  17 01 13.092 &  -30 06 55.195&         0.012   &0.012  &79.65   &Y/Y   & 93.07	 $^{+7.76}_{-7.35}$ /80.89	 $^{+6.45}_{-6.12}$  &6.926  & -  & qLMXB\\
$s16$ &  17 01 12.748 &  -30 06 55.438&         0.027   &0.026  &19.2    &Y/Y   & 12.69	 $^{+2.93}_{-2.54}$ /16.99	 $^{+3.00}_{-2.68}$  &6.075  & -  & CV \\
$s17$ &  17 01 14.452 &  -30 06 54.822&         0.067   &0.062  &17.51   &N/Y   & - /10.63	 $^{+2.48}_{-2.15}$  &22.111 & L  & \\
$s18$ &  17 01 13.313 &  -30 06 53.286&         0.034   &0.028  &17.64   &Y/Y   & 23.20	 $^{+3.98}_{-3.57}$ /14.82	 $^{+2.80}_{-2.49}$  &7.715  & -  & CV\\
$s19$ &  17 01 12.870 &  -30 06 53.256&         0.02    &0.02   &35.33   &Y/Y   & 41.57	 $^{+5.02}_{-4.65}$ /27.97	 $^{+3.82}_{-3.50}$  &3.964  & -  & CV\\
$s20$ &  17 01 15.062 &  -30 06 51.954&         0.073   &0.09   &13.99   &Y/Y   & 7.41	 $^{+2.54}_{-2.08}$ /2.75	 $^{+1.63}_{-1.19}$  &29.468 & S(1)  & \\
$s21$ &  17 01 13.220 &  -30 06 50.381&         0.01    &0.009  &112.83  &Y/Y   & 146.80 $^{+10.44}_{-9.96}$/140.40   	 $^{+8.25}_{-7.94}$  &5.542  & -  & BH\\
$s22$ &  17 01 12.701 &  -30 06 48.868&         0.058   &0.045  &13.78   &Y/Y   & 19.98	 $^{+4.08}_{-3.60}$ /12.10 	 $^{+2.57}_{-2.26}$  &1.38   & - & MSP\\
$s23$ &  17 01 10.403 &  -30 06 47.060&         0.056   &0.051  &11.04   &Y/Y   & 2.30	 $^{+1.45}_{-1.02}$ /3.49	 $^{+1.48}_{-1.16}$  &31.184 & -  & \\
$s24$ &  17 01 13.376 &  -30 06 45.834&         0.02    &0.016  &51.26   &Y/Y   & 46.54	 $^{+5.68}_{-5.26}$ /47.95 	 $^{+4.93}_{-4.61}$  &8.282  & -  & CV\\
$s25$ &  17 01 12.848 &  -30 06 45.443&         0.017   &0.017  &86.29   &Y/Y   & 110.80 $^{+8.61}_{-8.19}$ /119.90	 $^{+7.66}_{-7.37}$  &4.006  & S(1,2)  & CV\\
$s26$ &  17 01 13.365 &  -30 06 42.976&         0.05    &0.052  &6.42    &N/Y   & - /7.98	 $^{+2.15}_{-1.82}$  &9.751  & L*  & \\
$s27$ &  17 01 13.106 &  -30 06 42.218&         0.013   &0.012  &79.33   &Y/Y   & 83.68	 $^{+7.57}_{-7.15}$ /84.53 	 $^{+6.55}_{-6.24}$  &8.208  & -  & qLMXB\\
$s28$ &  17 01 11.625 &  -30 06 40.549&         0.028   &0.025  &49.18   &Y/Y   & 18.68	 $^{+3.71}_{-3.28}$ /30.54 	 $^{+3.95}_{-3.64}$  &17.625 & S(1)  & CV\\
$s29$ &  17 01 13.111 &  -30 06 38.536&         0.012   &0.01   &112.94  &Y/Y   & 108.80 $^{+8.32}_{-7.90}$ /114.4 	 $^{+7.47}_{-7.14}$  &11.592 & -  & CV\\
$s30$ &  17 01 13.726 &  -30 06 36.798&         0.03    &0.028  &30.93   &Y/Y   & 9.85	 $^{+2.71}_{-2.29}$ /21.44 	 $^{+3.35}_{-3.04}$  &17.412 & - & CV \\
$s31$ &  17 01 13.758 &  -30 06 32.393&         0.013   &0.012  &99.2    &Y/Y   & 100.10 $^{+8.04}_{-7.63}$ /69.04 	 $^{+5.82}_{-5.53}$  &21.070 & L  & qLMXB\\
$s32$ &  17 01 13.536 &  -30 06 31.956&         0.034   &0.034  &32.77   &Y/Y   & 23.92	 $^{+3.83}_{-3.48}$ /24.30 	 $^{+3.66}_{-3.33}$  &19.891 & S(1)  & CV\\
$s33$ &  17 01 12.905 &  -30 06 30.907&         0.11    &0.1    &5.72    &N/Y   & - /3.16	 $^{+1.62}_{-1.22}$  &18.543 & L*  & \\
$s34$ &  17 01 10.059 &  -30 06 28.134&         0.027   &0.025  &61.4    &Y/Y   & 29.44	 $^{+4.46}_{-4.07}$ /32.32 	 $^{+4.07}_{-3.75}$  &41.433 & S(1,2)  & CV\\
$s35$ &  17 01 13.456 &  -30 06 23.430&         0.133   &0.09   &8.56    &Y/Y   & 3.81	 $^{+1.88}_{-1.43}$ /2.63	 $^{+1.36}_{-1.03}$  &27.332 & -  & \\
$s36$ &  17 01 12.582 &  -30 06 22.201&         0.025   &0.024  &56.55   &Y/Y   & 28.75	 $^{+4.32}_{-3.92}$ /17.10 	 $^{+3.10}_{-2.77}$  &27.344 & - & qLMXB\\
$s37$ &  17 01 11.051 &  -30 06 21.265&         0.045   &0.051  &17.01   &Y/Y   & 6.92	 $^{+2.15}_{-1.78}$ /3.78	 $^{+1.55}_{-1.23}$  &36.141 & -  & \\
$s38$ &  17 01 13.171 &  -30 06 18.579&         0.079   &0.067  &14.66   &Y/Y   & 7.48	 $^{+2.25}_{-1.87}$ /2.72	 $^{+1.41}_{-1.10}$  &31.196 & -  & \\
$s39$ &  17 01 13.453 &  -30 05 59.364&         0.079   &0.101  &7.18    &Y/N   & 2.94	 $^{+1.54}_{-1.15}$ / -  &50.750 & L*  & \\
$s40$ &  17 01 13.532 &  -30 07 27.643&         0.104   &0.102  &9.89    &Y/Y   & 3.97	 $^{+1.85}_{-1.44}$ /2.14	 $^{+1.22}_{-0.95}$  &39.405 & -  & \\
$s41$ &  17 01 14.358 &  -30 06 48.361&         0.2     &0.108  &4.59    &N/Y   & - /1.29	 $^{+1.03}_{-0.70}$  &20.253 & L*  & \\
$s42$ &  17 01 12.567 &  -30 07 41.134&         0.088   &0.142  &5.28    &Y/N   & 4.09	 $^{+1.96}_{-1.50}$ / -  &8.799  & L*  & \\
$s43$ &  17 01 13.952 &  -30 06 20.635&         0.105   &0.135  &4.36    &Y/N   & 3.76	 $^{+2.94}_{-1.94}$ / -  &32.422 & L*  & AGN? \\
\hline
\end{tabular}
\begin{tablenotes}
      \centering 
      \item {

(1) Equatorial coordinates of sources in epoch J2000; (2) $1\sigma$ statistical errors of the X-ray positions; (3) the signal-noise ratios; (4) Flags for detection (Y) / non-detection (N) in individual observation; (5) Unabsorbed flux in 0.3-7~keV in individual observation; (6) Angular separation from the cluster centre; (7) Flags for indicating variability of each sources: S (short-term variable source) and L (long-term variable source). L* (long-term variable source candidates with non-detection in either obs1 or obs1) (8) Possible nature of the source identified in this work}
    \end{tablenotes}

\end{table*}

\section{Temporal Analysis}
Apart from the hardness and brightness in X-ray, we have also examined the temporal behaviours of the X-ray sources in M62. Since we have two observations with an epoch separation of 12 years, we are able to examine whether there is any long-term X-ray flux variabilities across $\sim12$~years as well as the short-term variabilities in each observation. 

\subsection{Searches for long-term variability}

For searching long-term X-ray flux variabilities of the sources in M62, we computed $S_{\rm flux}$ by the following equation:

\begin{equation}
S_{\text{flux}} = \frac{|F_{obs1}-F_{obs2}|}{\sqrt{{\sigma}_{{\text{F}}_{obs1}}^2+{\sigma}_{{\text{F}}_{obs2}}^2}},
\end{equation}

\noindent where $F_{obs1}/F_{obs2}$ and $\sigma_{{\text{F}}_{obs1}}/\sigma_{{\text{F}}_{obs2}}$ are the absorption-corrected X-ray flux and the corresponding errors measured in each epoch respectively (cf. column 8 in Table~1). 

Hence, the parameter $S_{\text{flux}}$ indicates the significance of the flux variation between two epochs. In this work, we consider an X-ray source has long-term variability if its $S_{\text{flux}}$ is larger than 3. With this criterion, only s31 which is considered as a qLMXB candidate is found to exhibit long-term X-ray flux variability. In the column 10 of Table~1, we indicate its variability by the flag "L". Together with the 11 sources which have non-detections in one of the two observations, which are indicated by the flags "L*" in the column 10 of Table~1, 12 long-term variable sources have been identified within the half-light radius of M62.

\subsection{Searches for short-term variability}

For searching the possible short-term variability in each observation, we used the CIAO script {\tt glvary} which is based on the Gregory-Loredo algorithm \citep{gl1992}. The algorithm breaks a single observation into multiple time bins and searches for any deviations among different bins. The variability is ranked by an index with a range from 0 to 10. In our work, we consider a source to be variable in a particular observation if its variability index computed by {\tt glvary} is $\geq6$. This corresponds to a probability of  $\geq90\%$ that the signal is genuinely variable.
 
With this pre-defined criterion, evidences of short-term X-ray flux variability have been found from 8 and 2 sources in obs1 and obs2 respectively. In the column 10 of Table~1, we indicate the short-term variable sources by the flag "S" with the numbers in the parentheses to specify in which observation(s) the variable signal is (are) detected. A majority of these variable sources are CV candidates. On the other hand, evidence for both long-term and short-term X-ray variabilities have been found in sources s5 and s14.

Sources s25 and s34 are found to have the largest short-term variability indices among all sources in both observations. In Figure~4, we compare their light curves obtained by each observation. The variabilities of these two CV candidates can be clearly seen. 

\begin{figure*}
    \begin{multicols}{2}
        \includegraphics[width=\linewidth]{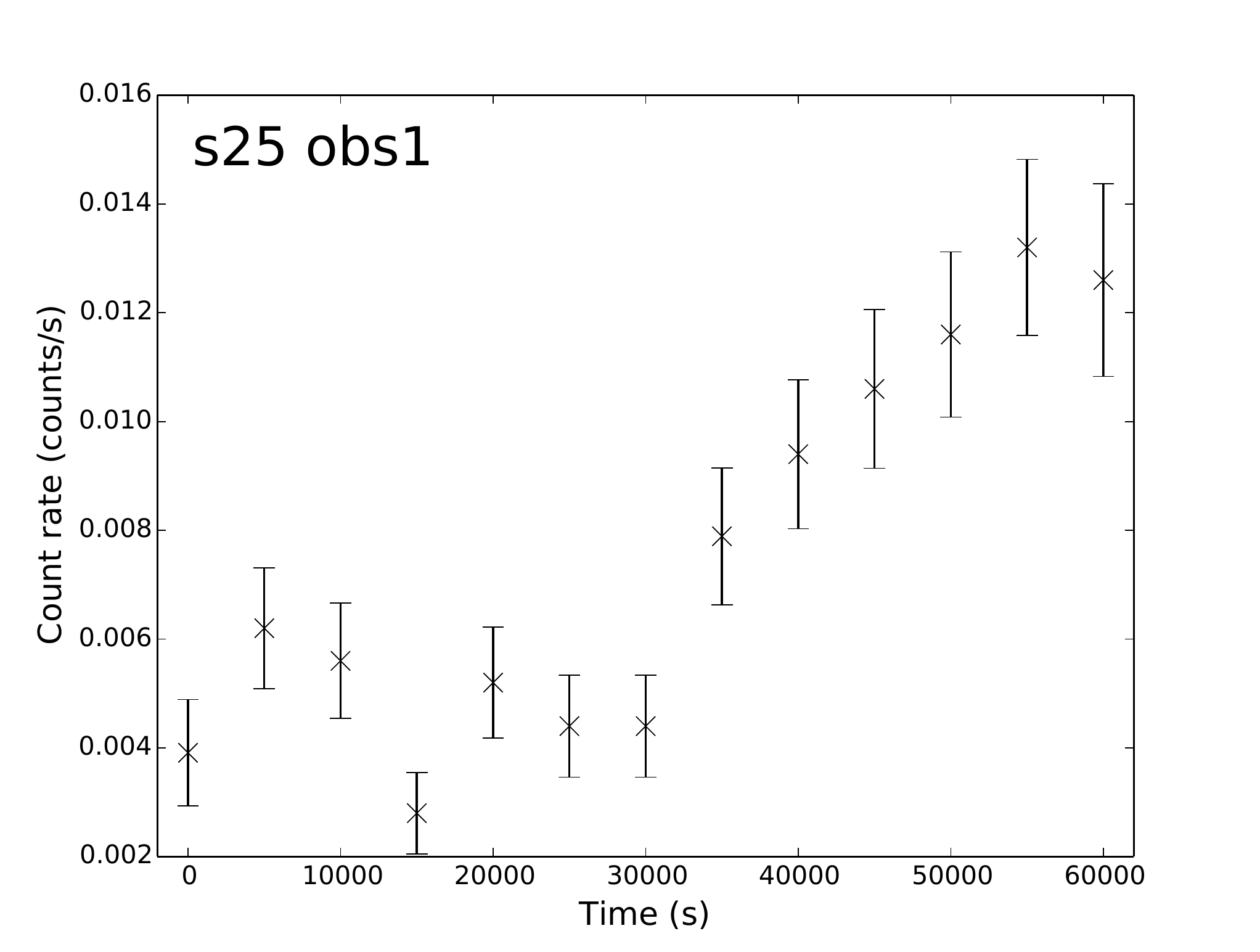}\par
        \includegraphics[width=\linewidth]{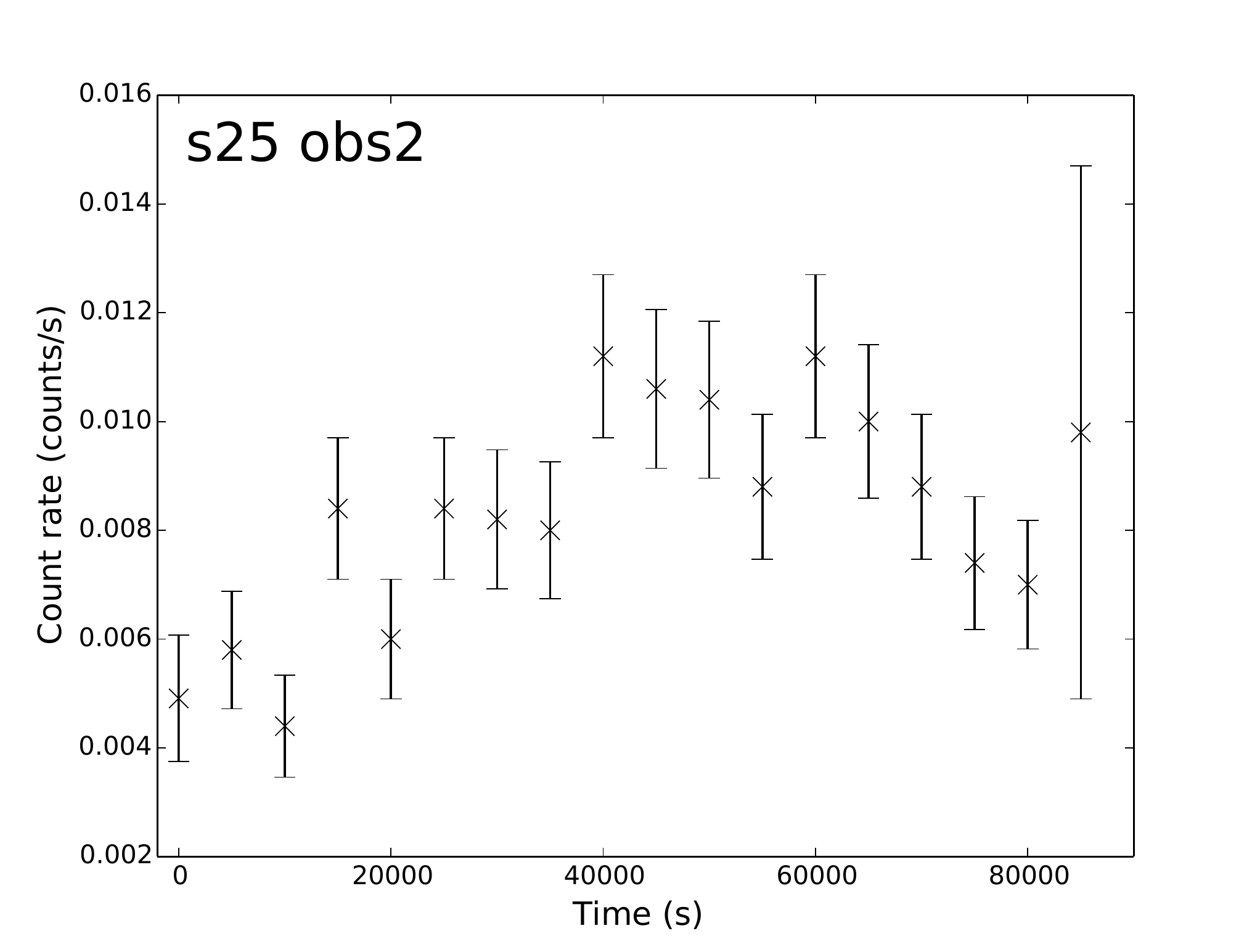}\par
        \end{multicols}
    \begin{multicols}{2}
        \includegraphics[width=\linewidth]{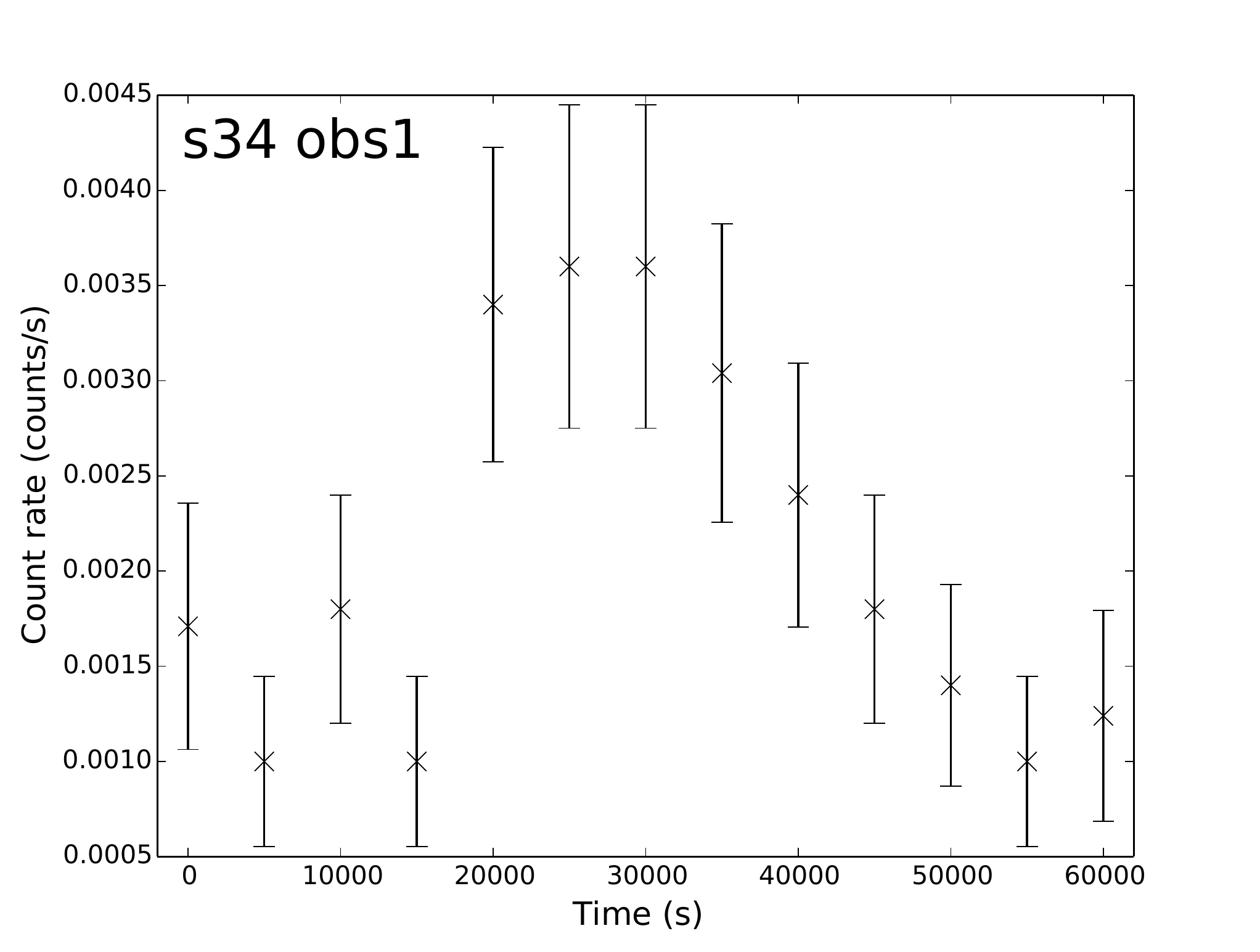}\par
        \includegraphics[width=\linewidth]{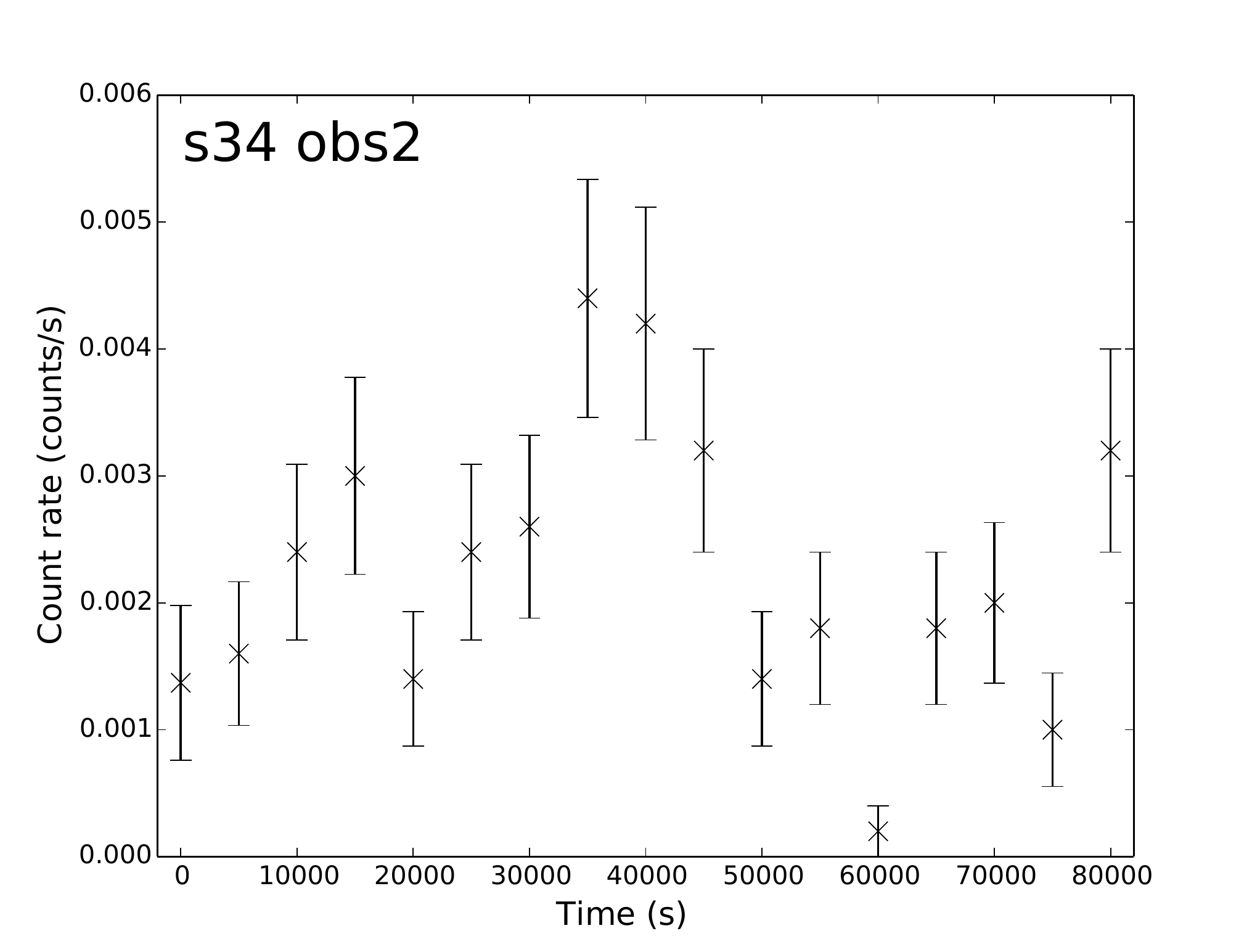}\par
        \end{multicols}
    \begin{multicols}{2}
        \includegraphics[width=\linewidth]{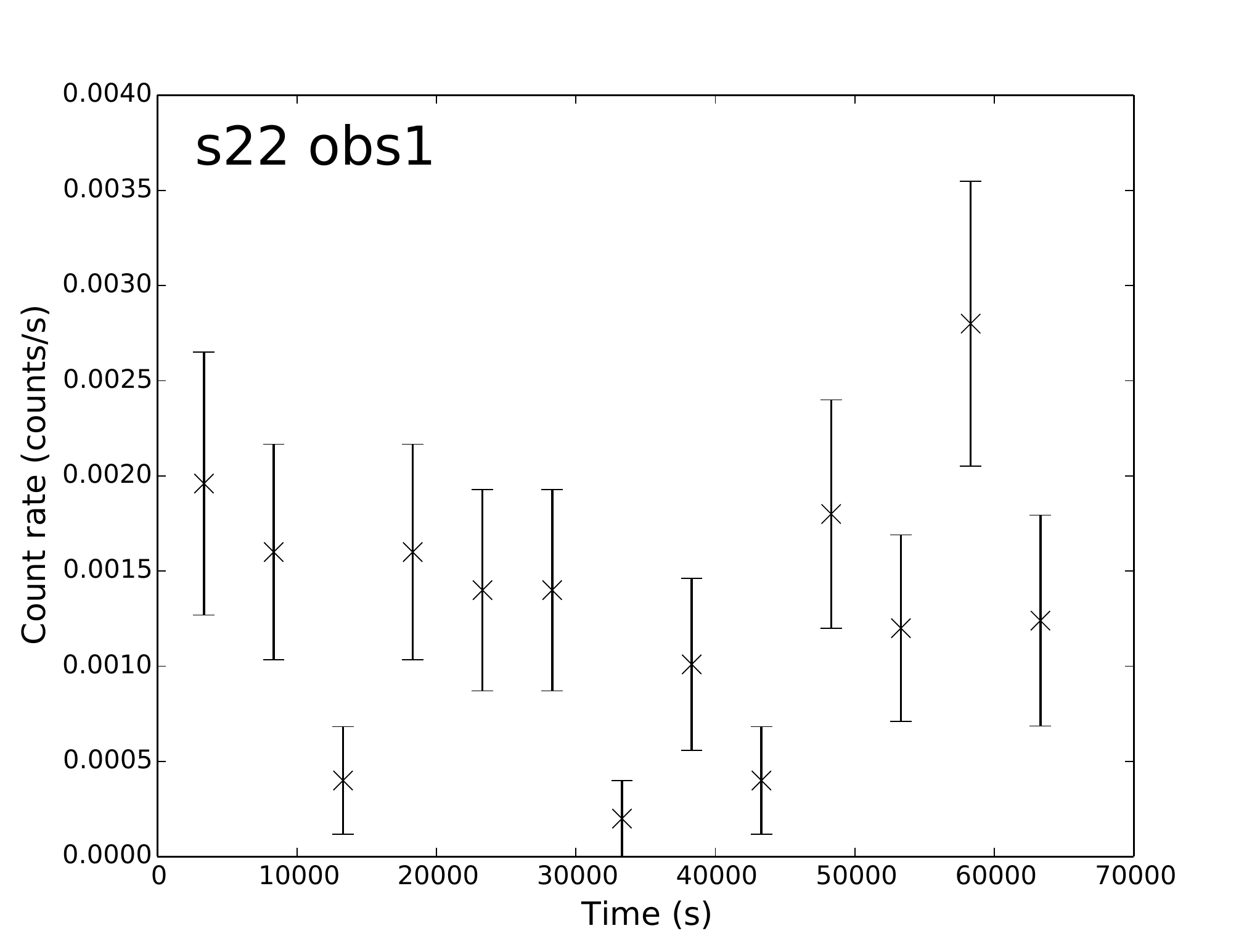}\par
        \includegraphics[width=\linewidth]{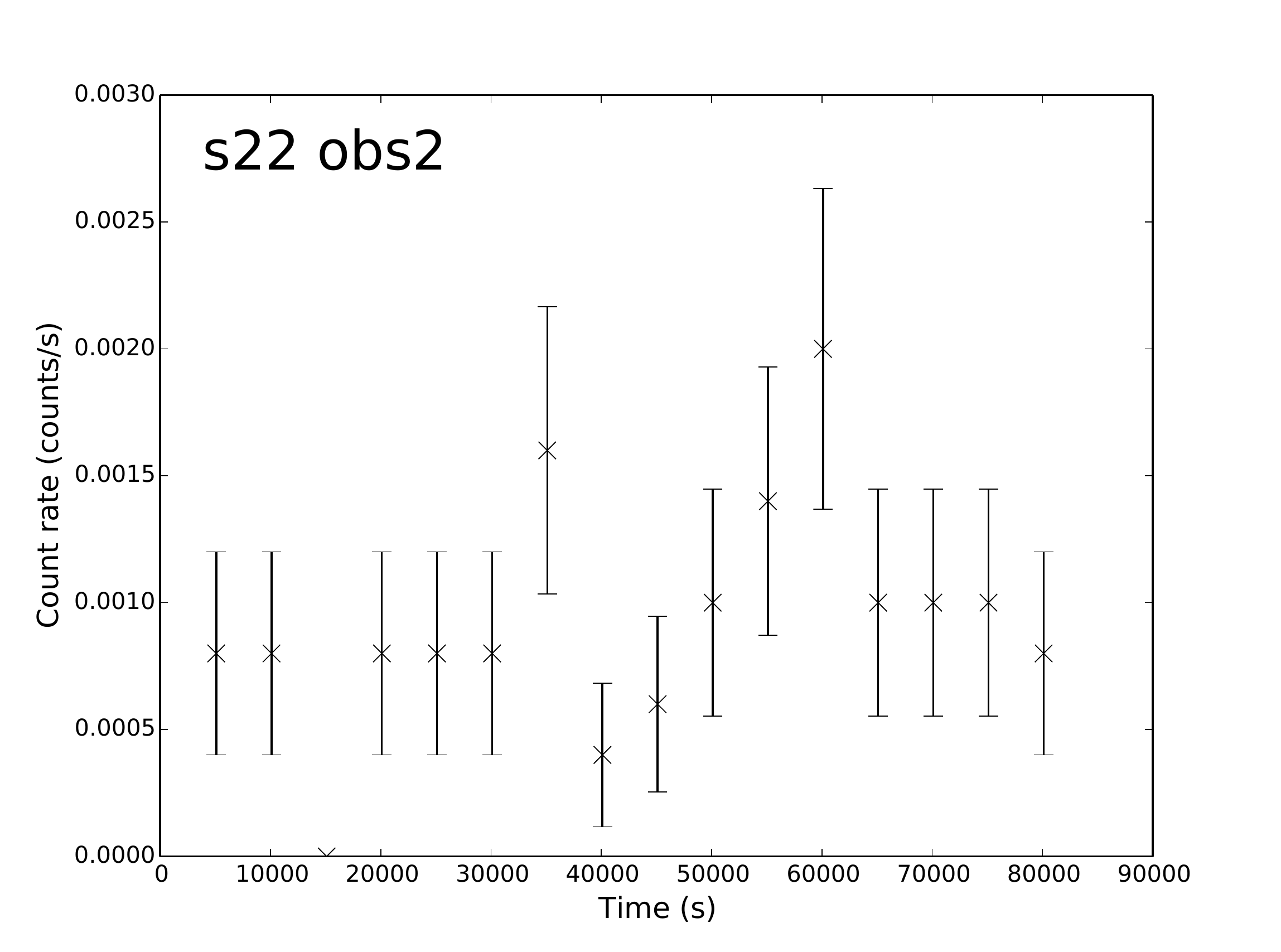}\par
        \end{multicols}
\caption{X-ray light curves of s25, s34 and s22 as observed by obs1 and obs2 in 0.3-7~keV. The origins of each plot are the starting time of the corresponding observation. The s25 and s34 are identified as CV candidates which have the most significant short-term variabilities found in both obs1 and obs2 (see Sec. 4.2). For s22, it is positionally coincident with PSR J1701-3006B which is a redback MSP (see Sec. 5.3).}
\end{figure*}

\section{Spectral Analysis}
For the bright sources have $\gtrsim150$ net counts collected by combining the data from both observations in the range of 0.3$-$7.0 keV, we proceeded to further investigate their emission nature by examining their X-ray spectra. This includes all the CV candidates, qLMXB candidates as well as the possible X-ray counterparts of a BH and a redback MSP identified in Section 3.3.  

We utilized the CIAO tool {\tt specextract} to extract the spectra and compute the response files. For the background spectra, they were sampled from the nearby source-free regions which are close to the corresponding X-ray source. All the spectral fittings were performed with XSPEC 12.6.0. The parameters of the best-fit models are summarized in Table~2 for all 18 selected sources. All quoted errors in this work are $1\sigma$ for 1 parameter of interest.

We have firstly examined the spectra of each source in each observation separately. By fitting each source spectrum with a simple absorbed power-law ($N_{H}$ fixed at $3.2\times10^{21}$~cm$^{-2}$), we found the spectral parameters inferred from both observations are consistent within the tolerance of their $1\sigma$ statistical uncertainties. In view of the lack of spectral variability, we proceeded to fit the spectra from obs1 and obs2 simultaneously in order to optimize the photon statistics. 

\subsection{Cataclysmic variable candidates}
We have first characterized the X-ray spectra of all CV candidates with a simple absorbed power-law (PL) model. The results are summarized in Table~2. Except for s30, the spectra of all these sources can be modeled by a simple PL reasonably well. The best-fit photon indices are in the range of $\Gamma\sim1-2$. Such phenomenological descriptions are comparable with those of magnetic CVs which typically exhibit a relatively hard spectrum with $\Gamma<2$ \citep[e.g.][]{hong2009,hong2012}.

The simultaneous spectral fit of s30 is not very satisfactory ($\chi^{2}=40.72$ for 26 d.o.f.). Visual inspection of its X-ray spectra obtained in obs1 and obs2 indicates there is a subtle difference in the hard band. Also, in calculating the long-term variability (Section 4.1), its $S_{\rm flux}$ is found to be 2.8 which is just below our predefined threshold. This shows that s30 is marginally variable. All these prompt us to examine each spectrum separately. For its spectrum in obs1, our best-fit model yields a photon index of $\Gamma=2.1^{+2.0}_{-1.2}$ and a normalization of $1.9\pm1.1\times10^{-6}$~photon~cm$^{-2}$~s$^{-1}$~keV$^{-1}$ at 1~keV with a goodness-of-fit of $\chi^{2}=6.29$ for 11 d.o.f.. On the other hand, the best-fit model in obs2 yield $\Gamma=2.0^{+0.7}_{-0.6}$, a normalization of $4.0\pm1.3\times10^{-6}$~photon~cm$^{-2}$~s$^{-1}$~keV$^{-1}$ at 1~keV with a goodness-of-fit of $\chi^{2}=11.93$ for 13 d.o.f..

Since the X-ray emission from a magnetic CV is likely originated from the shock-heated plasma above the surface of the white dwarf, we have also attempted to fit the X-ray spectra of all CV candidates with a more complicated model for the collisionally-ionized diffuse gas (XSPEC model: APEC). However, we found the spectral parameters cannot be properly constrained which can be ascribed to the relatively small photon statistics of these sources.

\subsection{Quiescent low-mass X-ray binaries candidates}
X-ray emission of a qLMXB can have multiple origins, such as the accretion disk and the heated neutron star surface. The X-ray spectrum resulting from accretion or any possibly pulsar activity can be phenomenologically determined by a PL model. Among four qLMXB candidates, s15, s31 and s36 can be well-described by a simple absorbed PL model, though their best-fit photon indices, $\Gamma>3$, are unphysically large. For s27, discrepancies between the PL model and the data are noticed (see Table~3).

We have also examined the spectra of these qLMXB candidates with an absorbed blackbody (BB) model which provides a phenomenological description of thermal X-rays from the neutron star surface. We found that such simple model can generally provide reasonable description for their X-ray emission. The best-fit models of these sources yield the temperature of $kT\sim0.2$~keV and a emitting region with a radius of $R_{\rm BB}\sim1.5$~km at distance of 6.8~kpc (see Table~3). 

To further examine their spectral properties, we have attempted to fit a composite PL+BB model to their spectra. For s27, s31 and s36, we found that their spectral parameters cannot be properly constrained in this composite model fitting and hence we will not further consider such model for these two sources. Among all the qLMXB candidates, only the PL+BB fit of s15 can yield a reasonable set of parameters ($\Gamma\sim1.8$, $kT\sim0.2$~keV and $R_{\rm BB}\sim1.8$~km). According to the best-fit parameters, the contribution of BB in the observed X-rays is $\sim3$ times that contributed by the PL component. Therefore, we conclude that the X-ray emission from all the qLMXBs in M62 are likely to be thermally-dominant.

\subsection{X-ray counterpart of redback millisecond pulsar PSR~J1701-3006B}
For the potential X-ray counterpart of redback MSP PSR~J1701-3006B, namely s22, we have examined whether its X-ray emission is thermally-dominant or non-thermally dominant by fitting its spectrum with both PL and BB models. The fitting results from both models are found to be satisfactory. For the PL fit, it yields a goodness-of-fit of $\chi^{2}=27.44$ with 28 d.o.f. and a photon index of $\Gamma\sim1.90$ which is comparable with the typical value for the general population of MSP in the Galactic field \citep{lee}. 

On the other hand, the BB fit yields a goodness-of-fit of $\chi^{2}=35.84$ with 28  d.o.f., a temperature of $kT\sim0.6$~keV and 
an emitting region with a radius of $R_{\rm BB}\sim50$~m. Although both model fittings are acceptable, the PL model apparently provides a better description of the data. Based on this, we suggest that the X-rays from s22  are likely to have a non-thermal origin. 

For redback MSPs, the major contribution for the X-ray emission is from the intrabinary shock which can result in variability or orbital modulation across the orbit \citep{hui2019}. Although evidence for neither long-term nor short flux variability has been found in s22, in view of its non-thermal nature suggested by the spectral analysis, we have examined its light curve to see if there is any subtle variability. In Figure~4, we have shown the X-ray light curves of s22 as obtained in obs1 and obs2. We have also examined if there X-ray modulation at the orbital period of PSR~J1701-3006B (i.e. $P_{b}\sim0.1$~days) in both obs1 and obs2 data. Nevertheless, we cannot find any conclusive evidence for X-ray orbital modulation in both observations.

\subsection{X-ray counterpart of black hole candidate M62-VLA1}
For the BH candidate s21, we found that its X-ray spectrum can be fitted with a simple absorbed PL model reasonably well ($\chi^{2}$=24.61 for 21 d.o.f.). The best-fit model yields a photon index of $\Gamma=2.6^{+0.3}_{-0.2}$ and a normalization of $3.6\pm0.5\times10^{-5}$~photon~cm$^{-2}$~s$^{-1}$~keV$^{-1}$ at 1~keV. This is consistent with the results reported by \cite{chomiuk} which are based on the analysis of obs1 data only. Our simultaneous spectral fit shows that s21 does not exhibit is any flux/spectral variability.

The lack of evidence for variability makes us question whether this source is indeed a BH. Furthermore, different from the other promising accreting BH candidates in GCs \citep{maccarone}, the X-ray luminosity of s21 is not high enough for us to rule out the other possible scenarios. \cite{chomiuk} have also mentioned that it is difficult to distinguish the BHs from neutron stars simply based on the X-ray spectrum.

The X-ray spectrum of s21 can also be well-fitted by a PL+BB model. The best-fit model yields a PL component of $\Gamma=2.2^{+0.4}_{-0.6}$ with
a normalization of $2.5^{+1.04}_{-1.17}\times10^{-5}$~photon~cm$^{-2}$~s$^{-1}$~keV$^{-1}$ at 1~keV and a BB component of $kT=0.14^{+0.06}_{-0.09}$~keV from an emitting region with a radius of 
$R_{\rm BB}\sim2.6$~km. We admit that the normalization of the BB component cannot be properly constrained. Despite that, our results are fully consistent with those reported by \cite{chomiuk}. Although $R_{\rm BB}$ is poorly constrained with the existing data, it is interesting to note that its value is similar to those inferred from the qLMXB candidates (see Table 3). Together with the non-detection of any X-ray flux variability, we question if this source is in fact a qLMXB instead of a BH.

\begin{table*}
 \captionsetup{justification=centering}
  \caption{Results of X-ray spectral fits of CV candidates in M62.}
   \begin{tabular}{cccc}
    \hline
    \multicolumn{4}{c}{\bf CV candidates in M62}\\
     \multicolumn{4}{c}{PL Model}\\
\hline
{Source} & {$\Gamma^{(1)}$} & {Norm$^{(2)}$} & {$\chi^{2}$/$\nu$}
\\
      &    & ($10^{-6}$ph/keV/cm$^2$/s)
\\
    \hline
$s06$  &2.06 $^{+0.36}_{-0.33}$ & 7.11$^{+1.72}_{-1.65}$ &  39.09/30 \\
$s11$  &1.07 $^{+0.29}_{-0.29}$ & 5.20$^{+1.59}_{-1.40}$ &  19.19/21 \\
$s16$  &1.38 $^{+0.71}_{-0.67}$ & 1.70$^{+1.05}_{-0.84}$ &  29.92/28 \\
$s18$  &1.61 $^{+0.69}_{-0.62}$ & 2.51$^{+1.28}_{-1.07}$ &  31.70/28 \\
$s19$  &1.66 $^{+0.57}_{-0.52}$ & 5.86$^{+2.24}_{-1.90}$ &   9.87/12 \\
$s24$  &1.29 $^{+0.36}_{-0.36}$ & 5.90$^{+1.88}_{-1.67}$ &  15.40/18 \\
$s25$  &1.36 $^{+0.20}_{-0.20}$ & 15.3$^{+2.81}_{-2.61}$ &  25.06/27 \\
$s28$  &1.55 $^{+0.43}_{-0.42}$ & 3.31$^{+1.24}_{-1.13}$ &  28.50/23 \\
$s29$  &1.34 $^{+0.19}_{-0.19}$ & 14.3$^{+2.53}_{-2.38}$ &  31.08/27 \\
$s30$  &1.97 $^{+0.69}_{-0.59}$ & 2.49$^{+1.22}_{-1.06}$ &  40.72/26 \\
$s32$  &1.24 $^{+0.50}_{-0.50}$ & 2.60$^{+1.13}_{-1.00}$ &  21.36/21 \\
$s34$  &0.98 $^{+0.41}_{-0.40}$ & 2.76$^{+1.17}_{-0.99}$ &  31.08/27 \\
    \hline
  \end{tabular}
\begin{tablenotes}\footnotesize

\item[*] $^{(1)}$ Photon index of the PL model.
\item[*] $^{(2)}$ Normalization of the PL model at 1 keV.
\end{tablenotes}
 \end{table*}

\begin{table*}
\centering
 \captionsetup{justification=centering}
  \caption{Results of spectral fits of qLMXB candidates in M62}
 \begin{tabular}{cccc|ccc}
    \hline
\multicolumn{7} {c}{\bf qLMXB candidates in M62}\\
    \hline
    \multicolumn{4} {c|}{PL model} & \multicolumn{3} {c}{BB model}\\
    \hline
{Source} & {$\Gamma$} & {Norm} & {$\chi^{2}$/$\nu$} & {$kT$} & R$_\text{BB}$$^{(1)}$   &  {$\chi^{2}$/$\nu$}
\\
 & & ($10^{-6}$ph/keV/cm$^2$/s) & & (keV) & (km) & \\
\hline
$s15$ &3.65 $^{+0.38}_{-0.38}$ & 25.8$^{+3.25}_{-3.27}$ &  31.58/24 &  0.198 $^{+0.021}_{-0.019}$    &1.46$^{+0.48}_{-0.37}$ & 24.34/24 \\
$s27$ &3.66 $^{+0.36}_{-0.35}$ & 28.8$^{+3.42}_{-3.45}$ &  37.74/23 &  0.212 $^{+0.021}_{-0.018}$    &1.28$^{+0.38}_{-0.30}$ & 28.82/23 \\
$s31$ &3.62 $^{+0.40}_{-0.40}$ & 27.0$^{+3.34}_{-3.36}$ &  23.39/21 &  0.204 $^{+0.021}_{-0.018}$    &1.37$^{+0.43}_{-0.35}$ & 13.26/21 \\
$s36$ &4.77 $^{+0.84}_{-0.77}$ & 7.62$^{+1.76}_{-1.78}$ &  16.53/17 &  0.156 $^{+0.034}_{-0.026}$    &1.57$^{+1.37}_{-1.57}$ & 21.47/17 \\
   \hline
\multicolumn{7} {c} {BB+PL model}\\
    \hline
{Source} & {$\Gamma$} & {Norm} &   \multicolumn{2}{c}{$kT$} & R$_\text{BB}$ &  {$\chi^{2}$/$\nu$}
\\
   & & ($10^{-6}$ph/keV/cm$^2$/s) &   \multicolumn{2}{c}{(keV)} & (km) & \\
   \hline
$s15$ &1.80 $^{+1.64}_{-1.80}$ & 4.53$^{+0.78}_{-4.36}$  & \multicolumn{2}{c}{0.174 $^{+0.099}_{-0.039}$}    & 1.88 $^{+1.50}_{-0.00}$ & 20.88/24 \\
    \hline

  \end{tabular}

\begin{tablenotes}\footnotesize
\item[*] $^{(1)}$ Radius of the emitting area in BB model with the assumption the source is at a distance of 6.8~kpc.

\end{tablenotes}

\end{table*}

\section{summary \& discussion} \label{sec:style}
We have performed a deep X-ray spectral imaging analysis of M62 by using two Chandra archival data with a total effective exposure of $\sim144$~ks. Within the half-light radius of the cluster, we have detected 43 point-like sources from the merged data (Figure~1 and Table~1). Based on the distribution of these sources in an X-ray colour-luminosity diagram (Figure~2 and Figure~3), 12 sources have been selected as CV candidates and 4 sources have been selected as qLMXB candidates. On the other hand, X-ray counterparts of 2 MSPs and a BH candidate are identified through positional coincidence. 

With an epoch separation of $\sim12$~years between these two observations, we have found evidence of long-term flux variability from 12 sources (see Table~1). On the other hand, short-term flux variabilities within an individual observation window have been found in 8 sources. Among them, s5 and s14 have exhibited both long-term and short-term flux variabilities.

For the bright sources, we have examined their X-ray spectra. There is no evidence of significant spectral variation for any sources between these two observations. The spectral fitting results are summarized in Section 5, Table 2 and Table 3. For the CV candidates, we found that their X-ray spectral steepness are comparable with those of magnetic CVs which can be characterized by a PL model with a photon index $\Gamma\sim1-2$. 

For the qLMXB candidates, we found that their X-ray emission are likely to be thermally-dominant. The best fit spectral parameter results of these 4 sources result from a simple BB model fitting are remarkably similar. They are characterized by a temperature of $kT\sim0.2$~keV and an emitting area with a radius of $R_{\rm BB}\sim1.5$~km. qLMXBs are progenitors of MSPs. Assuming the MSPs are active in these systems, the thermal emission can be originated from the hot polar caps with the radii $R_{\rm pc}$ as defined by the footprints of the last-open field lines: $R_{\rm pc}=R_{\rm NS}\sqrt{2\pi R_{\rm NS}/cP}$, where $R_{\rm NS}$, $P$ and $c$ are the radius of neutron star, rotation period and speed of light respectively. Assuming $R_{\rm BB}\sim R_{\rm pc}$ and $R_{\rm NS}\sim10$~km, this implies $P$ at the order of $\sim10$~ms which is comparable with the rotation period of MSPs. 

There are two major processes in forming an X-ray binary in a GC \citep{ivanova,heinke2010}. It can either form through dynamical interaction in the presence of frequent stellar encounters \citep{pooley2006,heinke2006} or it can have a primordial origin that dominating the formation of binaries in low density GCs. For investigating the origin of the X-ray binary population in M62, we compare its properties with two prototypical GCs --- 47 Tuc and Omega Cen. 

The low stellar density in Omega Cen implies a long relaxation time in the cluster. Its relaxation time is estimated to be 1.2$\times$10$^{10}$~yrs and 4.0$\times$10${^9}$~yr within its half-light radius and core radius respectively \citep{harris}. On the other hand, the time scales for the relaxation in 47 Tuc are much shorter which are estimated as $3.6\times10^{9}$~yrs and $6.9\times10^{7}$~yrs within its half-light radius and core radius respectively \citep{harris}. Such differences can be reflected in the distributions of their X-ray brightness as a function of angular distance from the cluster centres (see Figure~5). While the X-ray sources in Omega Cen are rather evenly distributed \citep[Figure 2 \& Figure~6 in][]{henleywillis}, the bright X-ray sources in 47 Tuc are somewhat concentrated towards the centre  \citep[Figure~7 in][]{heinke2005}. Since X-ray luminosity of an accretion-powered system scales with the mass of the accretor, the concentration of bright sources around the cluster centre can be a result of mass segregation. On the other hand, most of the faint unidentified sources lie outside the core which might suggest that the mass of these systems is lower. For example, it is likely that a fraction of these faint and hard X-ray sources are the counterparts of chromospherically active binaries. 

The relaxation times of M62 within its half-light radius and core radius are estimated to be $9.6\times10^{8}$~yrs and $7.9\times10^{7}$~yrs respectively \citep{harris}. This suggests its dynamical status is similar to that of 47 Tuc. Indeed, its distribution of X-ray luminosity as a function of angular distance from its centre shows that almost all the bright sources are located within 2$r_{c}$ from the centre (Figure 5).  

Apart from uncovering a more complete population of faint X-ray sources with deeper exposures, observations that cover a field-of-view larger than the half-light radius are also needed for further investigating the dynamical status of a GC.  
When objects sink to the centre of the cluster, a depletion region will be formed in an intermediate radial distance. This can be observed as a dip in the surface density distribution. Since massive objects sink to the centre faster, the dip of their distribution should locate further outward than that of less massive objects. Recently, \cite{cheng2019} have studied the radial distributions of the X-ray sources in 47~Tuc through a deep and wide-field investigation with archival {\it Chandra} data. While the dip for the radial surface density distribution of the faint sources ($L_{x}\lesssim5\times10^{30}$~erg/s) locates at $\sim100^{"}$ from the centre, the dip for the bright sources($L_{x}\gtrsim5\times10^{30}$~erg/s) locates further out at $\sim170^{"}$ \citep[see Fig.~5 in][]{cheng2019}. By fitting each distribution with a generalized King model, the average mass for the faint and bright X-ray sources are found to be $\sim1.16M_{\odot}$ and $\sim1.44M_{\odot}$ respectively \citep{cheng2019}. These are all consistent with the mass segregation of binaries in GCs. A similar analysis of M62 is encouraged to further explore the mass segregation effect of its X-ray sources.

The dynamical status of GCs can also be inferred by comparing their space densities of a specific type of binary. We choose CV for comparison as we have identified the largest number of their candidates among all classes in this study. By taking the distance of the outermost CV candidate from the centre as the radius of our region-of-interest, the space density of CV in M62 is estimated to be 0.85 pc${^{-3}}$. Applying the same scheme to the X-ray source catalog of 47 Tuc \citep{heinke2005}, the corresponding space density of CV is found to be 0.27 pc${^{-3}}$ which is a factor of $\sim3$ lower than that of M62. For Omega Cen, we found that its space density of CV is at the order of 0.016 pc${^{-3}}$ which is $\sim50$~times lower than that of M62.

\begin{figure*}
\includegraphics[width=\textwidth]{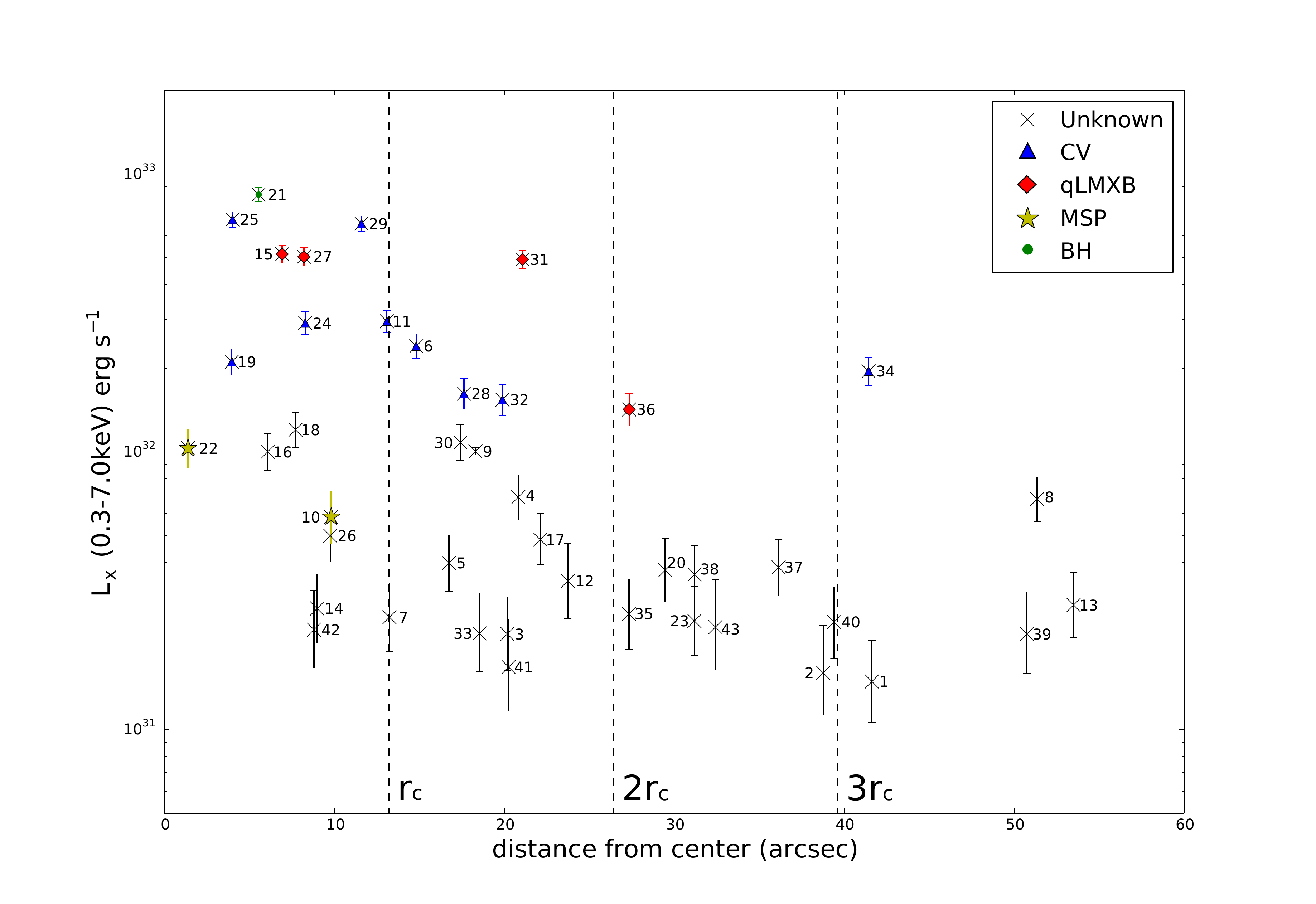}
\caption{X-ray luminosity of X-ray sources in M62 as a function of angular distance from the cluster centre. The dotted lines represent the angular distances of $r_{c}$, $2r_{c}$ and $3r_{c}$ from the cluster centre. Sources with different identified nature are represented by different symbols.}
\end{figure*}

We have also compared the X-ray luminosity distributions among these three GCs. For comparing with the source catalogs of 47~Tuc and Omega Cen \citep{heinke2005,henleywillis}, we rescale the X-ray luminosities of the sources in M62 in the energy range of 0.5-6~keV so as to be consistent with these catalogs. In the upper panels of Figure~6, we show the histograms (upper-left) and the empirical distribution functions (upper-right) of $\log L_{x}$ for the sources lie within the half-light radius of each GC and with $L_{x}\gtrsim7\times10^{30}$~erg/s (i.e. the faintest sources detected in M62). We also compare their distributions with larger threshold lumniosity ($L_{x}\gtrsim1.8\times10^{31}$~erg/s) in the lower-left and lower-right panels of Figure~6 for ensuring the completeness of source detection among all GCs.

It is interesting to note the differences among these three clusters, though the 2-sample Anderson-Darling tests show the statistical significance of their differences are all $<2\sigma$. In the sample with the threshold luminosity $L_{x}\sim7\times10^{30}$~erg/s, while the median X-ray luminosities in the samples of 47~Tuc and Omega~Cen are $\sim10^{31}$~erg/s, it is at the level of $\sim3\times10^{31}$~erg/s in M62. Furthermore, the fractions of the sources with $L_{x}\gtrsim10^{32}$~erg/s are found to be $\sim30\%$ in M62 and $\sim10\%$ in 47~Tuc/Omega~Cen. The larger fraction of the luminous X-ray sources in M62 is also found in the sample with the threshold luminosity of $L_{x}\gtrsim1.8\times10^{31}$~erg/s

The high fraction of luminous X-ray sources in M62 is consistent with its large X-ray emissivity \citep{cheng2018} which can be considered as a proxy of X-ray source abundance in a GC. This might shed light on its dynamical state. Stellar encounter rate $\Gamma_{c}$ is one standard parameter for comparing the dynamical status among GCs. $\Gamma_{c}$ of M62 is estimated to be $\sim2$ times of that in 47~Tuc \citep{arash2013}. This is in line with the fact that the X-ray emissivity of M62 is $\sim2$ times larger in M62 than 47 Tuc \citep{cheng2018}. 

Frequent stellar interactions in GCs can lead to orbital shrinkage of compact binaries, which is referred as collisional hardening \citep{Banerjee}. Together with magnetic braking and gravitational radiation, collisional hardening provides another means for shrinking the orbits of binaries in GCs and hence driving mass transfer. The characteristic orbital separation of the binaries in M62 is estimated to be $\sim50\%$ shorter than that in 47 Tuc (Table~5 in \cite{beccari}). This can possibly facilitate the mass transfer and therefore enhance the X-ray emission from the accretion-powered binaries. This is consistent with the conclusion deduced from a structural study which states that M62 is in a particularly active state in generating binaries through dynamical interactions \citep{beccari}. 

\begin{figure*}
    \begin{multicols}{2}
        \includegraphics[width=\linewidth]{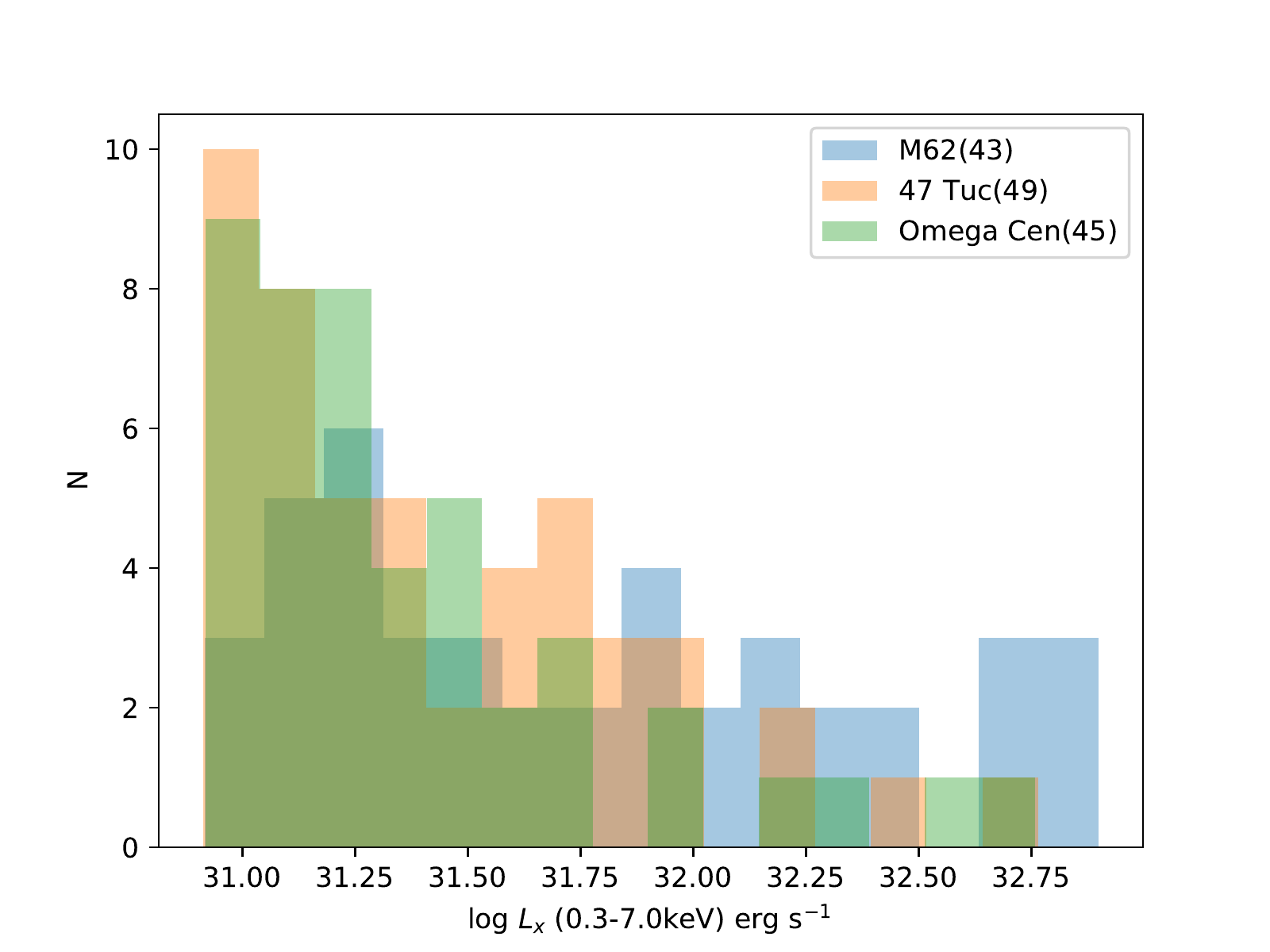}\par
        \includegraphics[width=\linewidth]{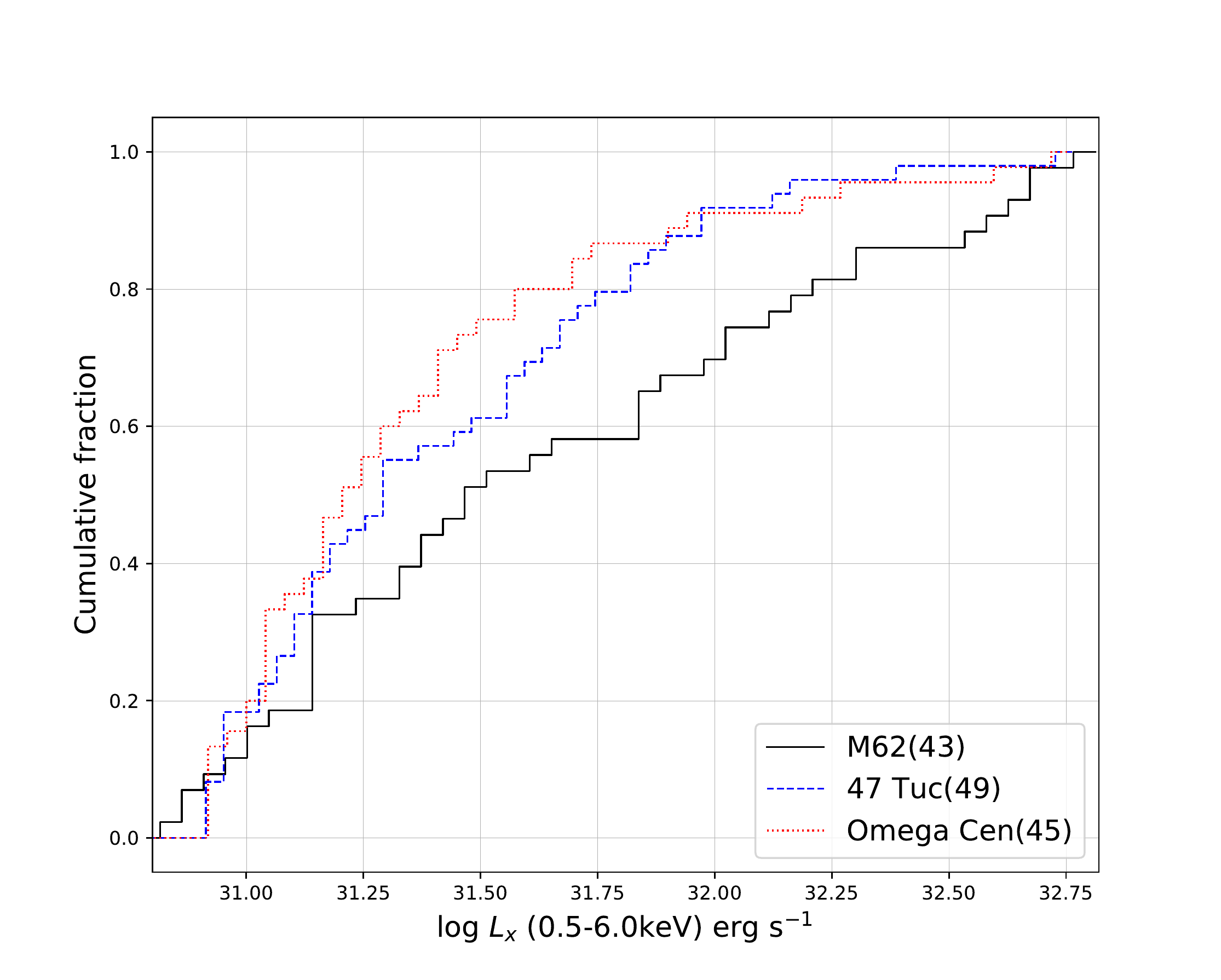}\par
        \end{multicols}
    \begin{multicols}{2}
        \includegraphics[width=\linewidth]{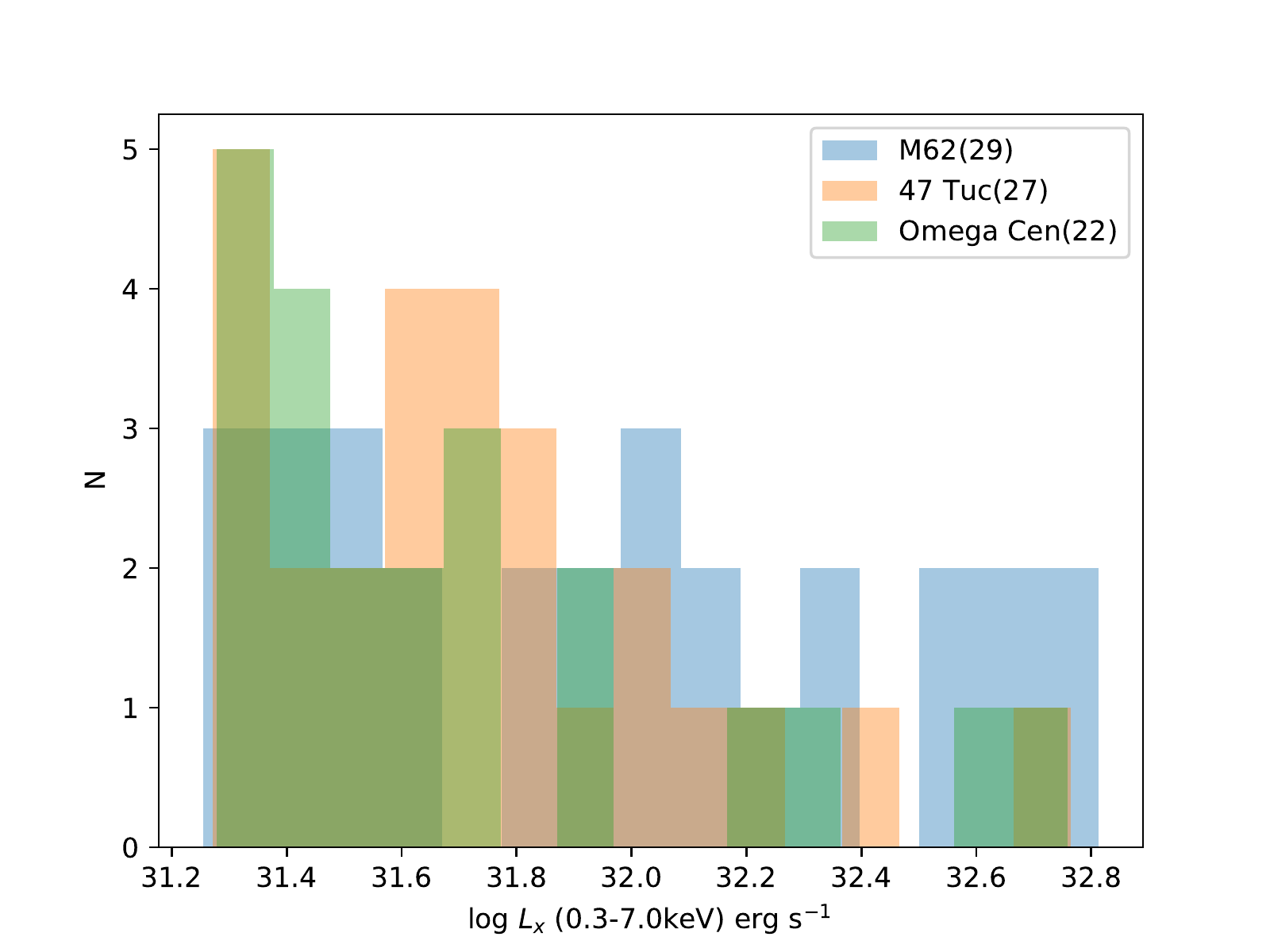}\par
        \includegraphics[width=\linewidth]{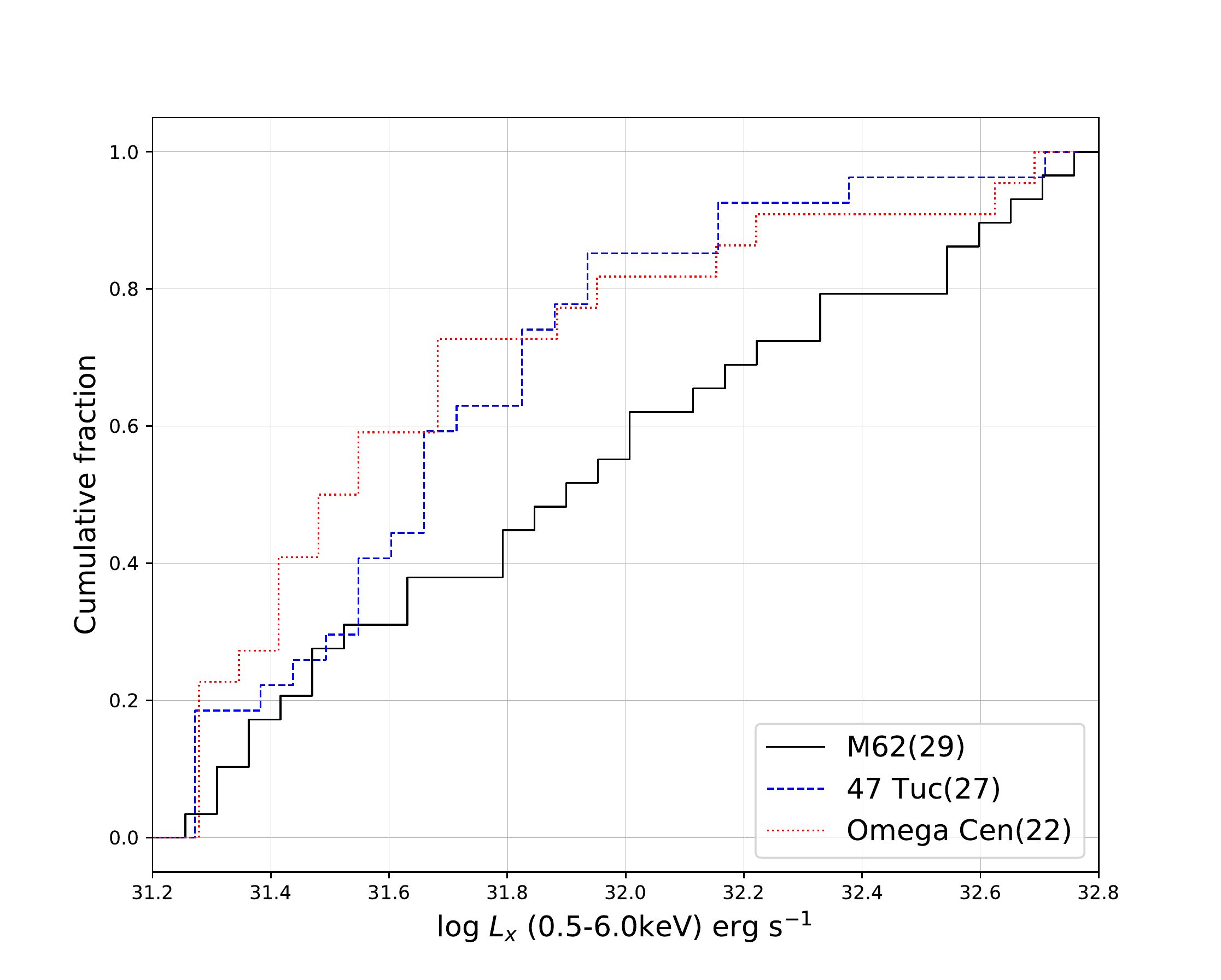}\par
        \end{multicols}
\caption{A comparison of histograms and empirical distribution functions for the X-luminosity of the sources within the half-light radii of three GCs: M62, 47~Tuc and Omega~Cen for the threshold luminosities of $L_{x}\gtrsim7\times10^{30}$~erg/s {\bf (upper panels)} and $L_{x}\gtrsim1.8\times10^{31}$~erg/s {\bf (lower panels)}. The sample size in each GC are given by the number in the legend.}
\end{figure*}

\section*{Acknowledgements}
KO is supported by National Research Foundation of Korea grant funded by the Korean Government (NRF-2019H1A2A1077058 Global Ph.D. Fellowship Program), 2016R1A5A1013277, 2019R1F1A1062071 and BK21 plus Chungnam National University; CYH is supported by the National Research Foundation of Korea through grant 2016R1A5A1013277 and 2019R1F1A1062071; KLL is supported by the Ministry of Science and Technology of Taiwan through grant 108-2112-M-007-025-MY3. AKHK is supported by the Ministry of Science and Technology of the Republic of China (Taiwan) through grants 105-2119-M-007-028-MY3 and 106-2628-M-007-005.

\section*{Data availability}
The data underlying this article were accessed from Chandra X-ray observatory [https://cda.harvard.edu/chaser/]. The derived data generated in this research will be shared on reasonable request to the corresponding author.


\bsp	
\label{lastpage}
\end{document}